\documentclass[prb,twocolumn,showpacs]{revtex4}
    \usepackage{epsfig,color}

\begin{document}
    \title{Magnetic forces and stationary electron flow \\ in three-terminal semiconductor quantum ring}
    \author{M.R. Poniedzialek } \affiliation{Faculty of  Physics and Applied
    Computer Science, AGH University of Science and Technology, al.
    Mickiewicza 30, 30-059 Krak\'ow, Poland}
    \author{B. Szafran} \affiliation{Faculty of Physics and Applied
    Computer Science, AGH University of Science and Technology, al.
    Mickiewicza 30, 30-059 Krak\'ow, Poland}

    \date{\today}

    \begin{abstract}
    We study stationary electron flow through a three-terminal quantum ring
    and describe effects due to  deflection of electron trajectories by
    classical magnetic forces. We demonstrate that generally at high magnetic field  ($B$)
    the current is guided by magnetic forces to follow a classical path which for $B>0$ leads via the left arm of the ring
     to the left output terminal. The transport to the left output terminal is blocked for narrow windows
     of magnetic field for which the interference within the ring leads to formation of wave functions that
     are only weakly coupled to the output channel wave functions. These interference conditions are accompanied by injection of the
     current to the right arm of the ring and by appearance of sharp peaks of the transfer probability to the right output terminal.
     We find that these peaks at high magnetic field are attenuated by thermal widening of the transport window. We also demonstrate
     that the interference conditions that lead to their appearance vanish when an elastic scattering within the ring is present.
     The clear effect of magnetic forces on the transfer probabilities disappears along with Aharonov-Bohm oscillations
     in a chaotic transport regime that is found for rings whose width is larger than the width of the channels.

    \end{abstract}
     \maketitle

    \section{Introduction}
Phase-coherent electron transport in mesoscopic\cite{but,web,timp} and nanoscale\cite{fuhrer,wgw,pedersen,keyser,muhle} rings  results in appearance
of Aharonov-Bohm\cite{ab} conductance oscillations in external magnetic field. These conductance oscillations are extensively studied in the context
of scanning gate spectroscopy,\cite{sgs} spin-orbit coupling for both electrons\cite{soe,lt} and holes,\cite{soh} Aharonov-Bohm interferometry\cite{qr} including electron self-interference,\cite{gus}
violation of Onsager symmetry,\cite{ihn} and magnetic forces.\cite{time,epl,strambini,poniedzialek}

The deflection of electron trajectories by magnetic forces in two-terminal quantum rings was previously studied by time-dependent wave packet simulations\cite{time} which indicated
that in presence of external perpendicular magnetic field the electron packet is preferentially injected into one of the arms of the ring which reduces
the Aharonov-Bohm interference of electron waves meeting near the exit to the output lead.
A time-dependent simulation was also used to describe the transport through a three-terminal quantum ring,\cite{epl}
which demonstrated that the Lorentz force -- besides the reduction of the Aharonov-Bohm oscillations at high field -- results in a distinct imbalance of the wave packet transfer probabilities to the two output leads. Such an imbalance of conductance of two output leads was indeed found in a recent experiment.\cite{strambini}

A three-terminal quantum ring is a basic element\cite{starypeeters}  for construction of ring arrays that are proposed
for implementation of quantum logic operations\cite{stare,peeters} using spin-orbit interactions.
In these structures\cite{stare,starypeeters,peeters} the direction of the charge current is determined by the
electron spin orientation. Magnetic forces\cite{strambini} can provide a mean of external control
of the current flow. 

The time-dependent simulations as previously performed for three-terminal rings\cite{epl,poniedzialek} are based on a relatively straightforward procedure
that indicates in a clear
way the electron trajectories across the nanostructure.  The charge transfer is a time-dependent process only in selected
experiments, cf. the single-electron injection into the quantum ring\cite{gus} realized according to the single-electron pump
technique based on the Coulomb blockade.\cite{poth} The standard experiments measure the current due to
the stationary electron flow at the Fermi level which is therefore of a more basic interest than the wave packet dynamics.
With the time-dependent approach one can in principle approach the monoenergetic time-independent limit increasing
the spatial spread of the wave packet in the initial condition but the latter is limited by necessarily finite size of the computational box.

The purpose of the present paper is to describe the effect of magnetic forces on electron transport through a three-terminal ring in Hamiltonian eigenstates.
We find that at high magnetic field the electron flow follows the
path determined by the Lorentz force -- one of the arms of the ring is selected by the current which leaves the ring to the nearest
output channel. However, exceptions to this rule are found for resonant interference conditions that block
the transport to the output channel that is preferred by magnetic forces. This blockade is accompanied by anomalous (nonclassical) injection of the
current to the ring
and by appearance of peaks of the transfer probability to the other output channel.
We study the thermal stability of this anomalous current injection,
the influence of the elastic scatterers for the resonant interference and effects of magnetic forces
in chaotic transport regime.
We also study oscillations of the current circulation which turn out to be more thermally stable than the oscillations of the transfer probabilities.
Orientation of the currents circulating inside the ring determines the sign of the magnetic dipole moment that they generate.
The magnetization oscillations due to the Aharonov-Bohm effect were so far measured for mesoscopic open quantum rings\cite{chch}
and for large ensembles of closed nanorings.\cite{fomin}

    \section{Theory}
    \subsection{Model system}
\begin{figure}[ht!]
     \hbox{\rotatebox{0}{\epsfxsize=70mm
                    \epsfbox[101 264 590 748] {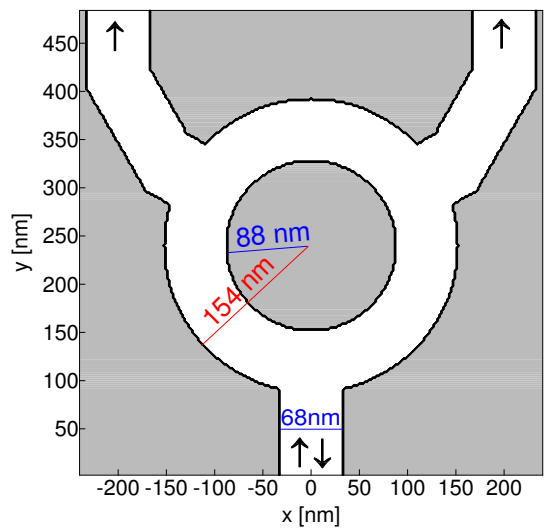}
                    \hfill}}
    \caption{Schematic drawing of the three-terminal ring.
    The confinement potential is zero inside the channels and 200 meV in the outside (grey area).
    The channel width is 68 nm, the inner and outer radii of the ring are 88 nm and 154 nm, respectively.
    The channel connected to the ring from below is the input lead.
     }
    \label{pot}
    \end{figure}

    The geometry of the studied system is depicted in Fig. \ref{pot}.
    The inner and outer radii of the ring are 88 nm and 154 nm, respectively. The channels
    are assumed $68$ nm wide. We treat the straight channel connected to the ring from below as the input terminal.
    The contacts to the input and the output channels are spaced
    by $120^\circ$ angles.
    The output channels are bent twice under the angle of $30^\circ$
    to acquire vertical orientation at the end of the computational box which
    allows for a uniform treatment of incoming and outgoing wave functions and currents (see below).

    We adopt a two-dimensional model and assume that the magnetic field is oriented perpendicular to the plane of confinement.
    We consider the electron Hamiltonian in form
    \begin{eqnarray}
    H&=&\left(-i\hbar\nabla+e{\bf A}({\bf
    r})\right)^2/{2m^*}+V(x,y) 
    \label{xx}
    \end{eqnarray}
    where $V(x,y)$ is the confinement potential -- assumed zero within the channels (white area in Fig. \ref{pot}) and $V_0=200$ meV
    in the outside (the grey area in Fig. \ref{pot}). The potential offset $V_0$ corresponds to channels made of GaAs
    embedded in an Al$_{0.45}$Ga$_{0.55}$As matrix. In Eq. (\ref{xx}) $-e$ is the electron charge ($e>0$) and $m^*=0.067m_0$ is the GaAs electron band effective mass.

    \subsection{Hamiltonian discretization}
    For the description of the stationary charge transport through the system we need to determine the Hamiltonian (\ref{xx}) eigenfunctions for the electron coming from the input channel.
    We employ the finite difference approach with a square computational box of side length
    482 nm (see Fig. \ref{pot}) on a grid  of $241\times 241$ points
    with mesh spacings $\Delta x=\Delta y=2$ nm. The results presented below are unaffected when one enlarges the computational box to cover a larger part of the input and the output channels. We use the Wilson\cite{wilson} type of discretization
    of the kinetic energy operator in a version adapted by Governale and Ungarelli \cite{governale} for
    semiconductor nanostructures. The discretization is consistent with the original Hamiltonian [tends
    to it in the $\Delta x=0$ limit] and gauge-invariant [accounts for the gauge transformation ${\bf A}\rightarrow {\bf A}+\nabla \chi({\bf r})$ inducing wave function phase change
    $\Psi({\bf r})\rightarrow \exp\left(-\frac{ie}{\hbar} \chi({\bf r})\right)\Psi({\bf r})$].
    The kinetic energy operator\cite{governale} acting on a wave function defined on a mesh
    yields
    \begin{eqnarray}
    &&\frac{1}{2m^*}\left({\bf p}+e{\bf A}\right)^2\Psi_{\mu ,\nu}=\nonumber \\ &&\frac{\hbar^2}{2m^*\Delta x^2} \left(4\Psi_{\mu,\nu}-C_y \Psi_{\mu,\nu-1}-C^*_y\Psi_{\mu,\nu+1}\nonumber \right.\\ && \left.-C_x \Psi_{\mu-1,\nu}-C^*_x\Psi_{\mu+1,\nu}\right),\label{governale}
    \end{eqnarray}
    where $\Psi_{\mu,\nu}=\Psi(x_\mu,y_\nu)$,  $C_y=\exp\left[-i\frac{e}{\hbar}\Delta x A_y \right]$, and $C_x=\exp\left[-i\frac{e}{\hbar}\Delta x A_x \right].$
    We apply the Lorentz gauge ${\bf A}=(A_x,A_y,0)=(0,Bx,0)$, for which
    the mesh Hamiltonian  eigenequation reads
    \begin{eqnarray}
    H\Psi_{\mu,\nu}&=&\frac{\hbar^2}{2m^*\Delta x^2} \left(4\Psi_{\mu,\nu}-C_y \Psi_{\mu,\nu-1}-C^*_y\Psi_{\mu,\nu+1}\right.\nonumber \\ && \left. - \Psi_{\mu-1,\nu}-\Psi_{\mu+1,\nu}\right)+V_{\mu,\nu}\Psi_{\mu,\nu}\nonumber \\ &=&E\Psi_{\mu,\nu}. \label{new}
    \end{eqnarray}
    We find the energy  $E$ by solution of the  boundary problem in the incoming lead (see the next section), then
     Eq. (\ref{new}) is solved as a system of linear equations.

    \subsection{Boundary conditions}
    The confinement potential in both the input and the output channels depends only on the $x$ coordinate. The chosen gauge
    allows for separation of the $x$ and $y$ coordinates in
    the Hamiltonian eigenfunctions
    \begin{equation}
    \Psi(x,y)=\exp(iky)\psi_{n}^k(x), \label{psi}
    \end{equation}
    with the wave vector $k$.
    In the absence of the magnetic field
    the $n$th Hamiltonian eigenstate of a $w=68$ nm wide channel has the energy $E_n=n^2\pi^2/2m^*w^2=1.21n^2$ meV.
    We consider the transport limited to the lowest $n=1$ subband and skip the subscript $n$ in the
    following. Only the electrons with wave vector exceeding $k\simeq 0.08 $ nm$^{-1}$
    have enough energy to be scattered to higher subbands. We restrict our discussion to lower values of $k$.
    According to the Landauer-B\"uttiker approach\cite{lba} in the single subband transport the conductance $G$ is simply proportional to  the transfer probability $G=\frac{2e^2}{h}T$.


    We first determine the boundary conditions for the incoming lead. We assume $\nu=1$ (lowest row of the mesh in the computational box)
    and plug
    \begin{equation}
     \Psi_{\mu,\nu \pm 1}=\exp(\pm i k\Delta x)\psi^k_{\mu}
    \end{equation}
    into Eq. (\ref{new}) to obtain one-dimensional eigenequation
    \begin{eqnarray}
    &&\frac{\hbar^2}{2m^*\Delta x^2} \left(2\psi^k_{\mu}- \psi^k_{\mu-1}-\psi^k_{\mu+1}\right)\nonumber \\
    &&+\frac{\hbar^2}{m^*\Delta x^2} \left(1-\cos(k\Delta x + \frac{e}{\hbar}Bx\Delta x)\right) \psi^k_{\mu}
    \nonumber \\  &&
    +V_{\mu}\psi^k_{\mu}=E\psi^k_{\mu}.\label{new2}
    \end{eqnarray}
    Eq. (\ref{new2}) provides the energy $E$ that is used in the main equation (\ref{new})
    as well as eigenfunctions corresponding to incident ($k>0$) and reflected ($k<0$) electrons
    that are used for setting the Dirichlet boundary condition for Eq. (\ref{new}) at the bottom of the computational box
    \begin{equation}\Psi_{\mu,\nu=1}=c_k\psi^k_\mu+c_{-k}\psi^{-k}_\mu,\label{dirichlet}\end{equation} where the amplitudes of the incident $c_k$
    and reflected $c_{-k}$ wave functions  are determined in a manner described in Section \ref{subs}.

    Boundary condition (\ref{dirichlet})
    guarantees that the energy density $H\Psi(x,y)/\Psi(x,y)=E$ in the incoming lead and within the ring
    as found from Eq. (\ref{new}) are equal.
    In order to match the energy density inside the ring
    the   wave vectors in the output channels [see Eq. (\ref{psi})] in nonzero $B$ must be different than $k$. Within the channels the confinement potential is zero,
    therefore equal energy density means equal kinetic energy density.
    The kinetic energy operator is proportional to the square of the  kinetic momentum $\Pi^2=(p+e{\bf A})^2=\Pi_x^2+\Pi_y^2=-\hbar^2 \frac{\partial ^2 }{\partial x^2}+\left(-i\hbar \frac{\partial  }{\partial y}+eBx\right)^2$.
    Since the input and the output channels have the same width, the energy density is matched for the
    wave vectors in the left $k_l$ and right $k_r$ output leads related to the wave vector of the incoming lead $k$
    as $k_l=k-\frac{eB}{\hbar} x_l$, and $k_r=k-\frac{eB}{\hbar}x_r$, where $x_l$ and $x_r$ are positions of the axes of the left and right output leads
    ($x_r=-x_l=200$ nm).
    Accordingly, for the boundary condition at the top of the computational box we use
    \begin{equation}
    \Psi_{\mu,\nu+1}=\Psi_{\mu,\nu}\exp(ik'\Delta x), \label{bcup}
    \end{equation}
    where $k'=k_l$ for $x<0$ and $k'=k_r$ for $x>0$.
    This condition is introduced into Eq. (\ref{new}) for the top end of the computational box ($\nu=241$).


    On the left and right edges of the computational box we introduce an infinite potential barrier which amounts
    in putting $\Psi_{\mu-1,\nu}=0$ or $\Psi_{\mu+1,\nu}=0$ in Eq. (\ref{new}) for a mesh points at the left and right ends of the box,
    respectively.

    \begin{figure}[ht!]
     \hbox{\rotatebox{0}{\epsfxsize=60mm
                    \epsfbox[20 145 564 622] {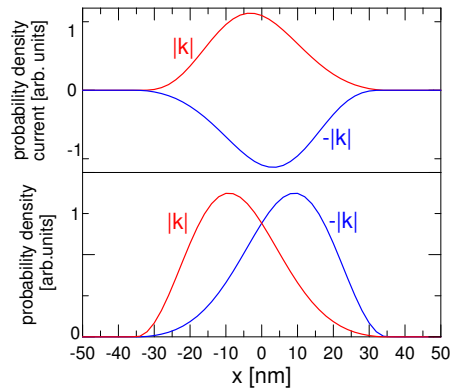}
                    \hfill}}
    \caption{Probability density (lower panel) and probability density current (upper panel)
    across the incoming lead for the incident $\Psi^{|k|}(x)$ and backscattered $\Psi^{-|k|}(x)$ electron eigenfunctions for $|k|=0.05$ nm$^{-1}$
    at $B=1$ T. }
    \label{prondy}
    \end{figure}

    \begin{figure}[ht!]
     \hbox{\rotatebox{0}{\epsfxsize=60mm
                    \epsfbox[20 66 564 600] {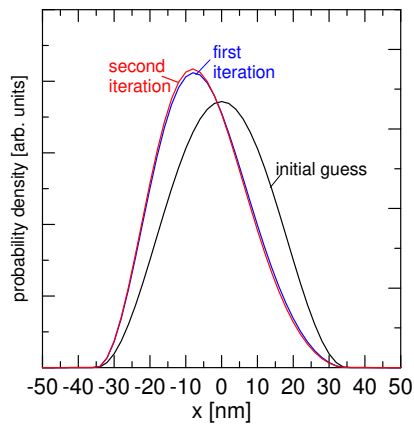}
                    \hfill}}
    \caption{The probability density in the incoming lead $c_k\Psi^k(x)+c_{-k}\Psi^{-k}(x)$ in the initial guess $c_k=c_{-k}$
    and in the subsequent iterations of the self-consistent procedure (see text). Parameters are same as in Fig. \ref{prondy}. }
    \label{incoming}
    \end{figure}

    \subsection{Backscattering probability}
    The vertical component of the probability density current in the incoming lead
    \begin{equation}
    j(x)=\frac{\hbar}{m^*}\Im(\Psi^*\frac{\partial \Psi}{\partial y})+\frac{e}{m^*}A_y, \label{gcf}
    \end{equation}
    is a superposition $j(x)=j^{k}(x)+j^{-k}(x)$ of the incident current
    \begin{equation}
    j^{k}(x)=\frac{\hbar }{m^*}|c_k|^2|\psi^k(x)|^2(\hbar k+eBx) \label{ji}
    \end{equation}
    and the backscattered one
    \begin{equation}
    j^{-k}(x)=\frac{\hbar }{m^*}|c_{-k}|^2|\psi^{-k}(x)|^2(-\hbar k+eBx).
    \end{equation}

    Figure \ref{prondy} shows the probability density and probability density current across the incoming lead
    for the incident ($k$) and reflected ($-k$) waves  for $k=0.05$ nm$^{-1}$ and
    $B=1$ T. Probability densities are shifted from the axis of the lead to the left with respect
    to the direction of the current flow in consistence with the Lorentz force orientation.
     The basckattering probability is evaluated as a ratio
    of the current fluxes integrated across the input channel
    \begin{equation}
    R=\frac{\int dx j^{-k}(x)}{\int dx j^{k}(x)}.
    \end{equation}
    For the axis of the incoming lead $x=0$, the solutions of the eigenequation (\ref{new2})
    with opposite $k$ are related as
     $\psi^k(x)=\psi^{-k}(-x)$ which implies that
     1) the backscattering probability is simply
    \begin{equation}
    R=\left|\frac{c_{-k}}{c_{k}}\right|^2,
    \end{equation}
    and
    2)  both the incident and reflected wave function correspond to the same
    average value of $\Pi_y^2$.

    Solution of the system of equations (\ref{new2}) gives the wave function in the entire system. Now our task is to extract
    $c_{\pm k}$, i.e. the contributions of the incident and backscattered wave functions.
    For that purpose we consider two points in the incoming lead
    near the bottom of the computational box. We typically take two lowest points of the axis of the lead
    ($\mu=120,\nu=1$) and ($\mu=120,\nu=2$), the results are not affected by a specific choice of these points.
    Wave function for $\nu=1$ is given by Eq. (\ref{dirichlet})
    and for $\nu=2$ we have
    \begin{equation}
    \Psi^k_{\mu,\nu=2}=c_k\psi^k_\mu\exp(ik\Delta x)+c_{-k}\psi^{-k}_\mu\exp(-ik\Delta x). \label{other}
    \end{equation}
    The eigenfunctions $\psi^k_{\mu}$, $\psi^{-k}_{\mu}$ are determined from Eq. (\ref{new2}). Formulas
    ($\ref{dirichlet}$) and ($\ref{other}$) form the system of equations  for $c_k$ and $c_{-k}$.

    \subsection{Self-consistence for the amplitudes of the incident and reflected wave functions}
    \label{subs}
    For nonzero $B$ the Hamiltonian ($\ref{new2}$) depends
    on the sign of the wave vector, and the eigenfunctions for $\pm k$ are different.
    We need to assume some initial values for $c_k$ and $c_{-k}$
    to set the boundary condition (\ref{dirichlet}) for the system of equations ($\ref{new}$).
    Solution of Eq. (\ref{new})
    gives the wave function in the entire computational box, including the incoming lead, of which $c_k$ and $c_{-k}$ can be extracted.
    The procedure to determine  $c_k$ and $c_{-k}$ is performed in a self-consistent iteration with $c_k=c_{-k}=\frac{1}{\sqrt{2}}$ assumed
    as the initial guess.  The iteration converges quite fast. Fig. \ref{incoming} shows the charge density
    across the incoming lead for $k=0.05$ nm$^{-1}$ at $B=1$ T. The final result differs considerably from the initial guess
    but the results of the second iteration only slightly differs from the first one. For parameters applied in Fig. \ref{incoming}
    the result for the backscattering probability $R=|c_{-k}|^2/|c_k|^2$ converges from 1 (for the initial guess) to 0.021.
    Naturally, the iteration affects the results in the entire computational box. 

     Note, that by the initial guess $c_k=c_{-k}$ one assumes that the wave function in the incoming lead is symmetric with respect to its axis.
        For nonzero $B$ this is the case only when backscattering probability reaches 100\%.

    \subsection{Transfer probabilities to the left and right output channels}
    We need to separate the electron transfer probability to the left $T_l$ and right $T_r$ leads of the total
    transfer probability $T=1-R$. For that purpose we calculate the probability currents in the left and right output lead
    at the top of the computational box. Formula (\ref{gcf}) with the boundary condition (\ref{bcup}) gives
    \begin{equation}
    j_l(x)=\frac{\hbar}{m^*}|\Psi(x,y')|^2\left(\hbar k_l+eBx)\right) \label{lr}
    \end{equation}
    for the left lead, and
    \begin{equation}
    j_r(x)=\frac{\hbar}{m^*}|\Psi(x,y')|^2\left(\hbar k_r+eBx)\right) \label{rr}
    \end{equation}
    for the right one, where $y'$ is the coordinate of the top of the computational box.
    We integrate the current fluxes on the left and right sides of the box $J_l=\int_{-120 \Delta x}^0 dx j_l(x)$ and $J_r=\int_0^{120 \Delta x} dx j_r(x)$. The transfer probability to the left and right channels is then calculated as $T_l=T\frac{J_l}{J_l+J_r}$
    and $T_r=T\frac{J_r}{J_l+J_r}$.  
    \subsection{Time dependent simulations}
    For the interpretation of the results it is useful to consider also the solution of the time-dependent Schr\"odinger equation,
    $i\hbar \partial \Psi/\partial t=H\Psi$.
    For the initial condition we use a Gaussian wave function entirely localized in the
    input lead
    \begin{eqnarray}
    \Psi(x,y,t=0)=\frac{\Delta k^{1/2}}{(2 \pi)^{1/4}}
    \psi_k(x) e^{-\frac{\Delta k^2}{4}(y-Y)^2+Iqy},
    \label{ic}
    \end{eqnarray}
    where $Y$ lies far enough below the ring in the incoming lead. Probability density
    of the initial condition in the wave vector space
    is
    \begin{equation}
    |\Psi(k)|^2=C\exp\left(-2(k-q)^2/\Delta k^2\right). \label{ick}
    \end{equation}
    The time-dependent calculations are performed using the Crank-Nicolson scheme with a time step of 0.3 fs.
     We use the finite difference Hamiltonian (\ref{new}) with the same mesh spacings as in the time independent calculation,
     but with radically enlarged computational box. For the time dependent simulation
    the computational box that we use covers as much as 12 $\mu$m of the input and the output channels. Computational box of this large size was necessary
    for setting the initial condition for a nearly monoenergetic wave packet.

    \begin{figure}[ht!]
    \begin{tabular}{ll}
     a) &  \epsfxsize=70mm \epsfbox[30 120 570 423] {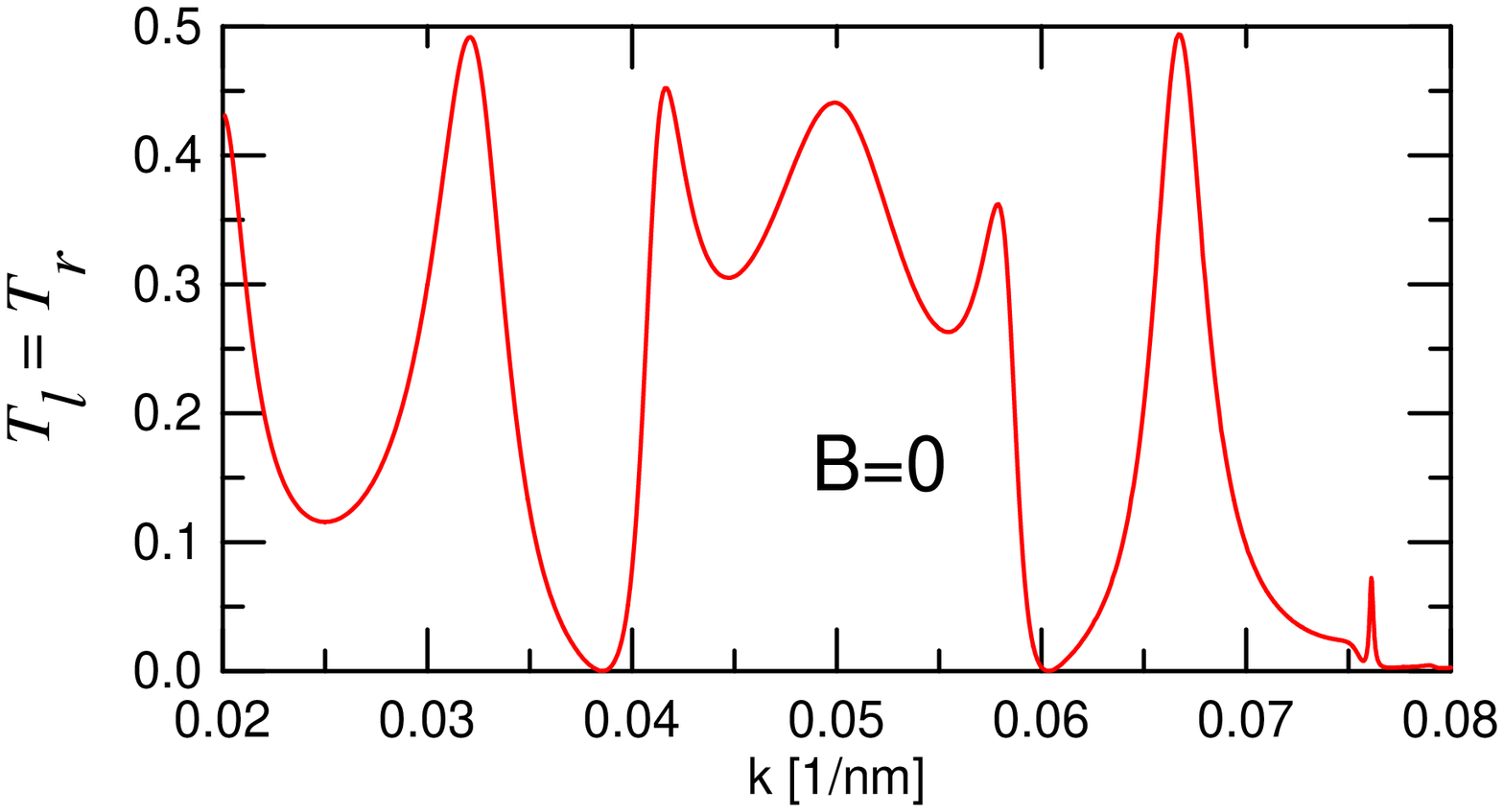} \\
     b) &   \epsfxsize=70mm  \epsfbox[30 120 570 423] {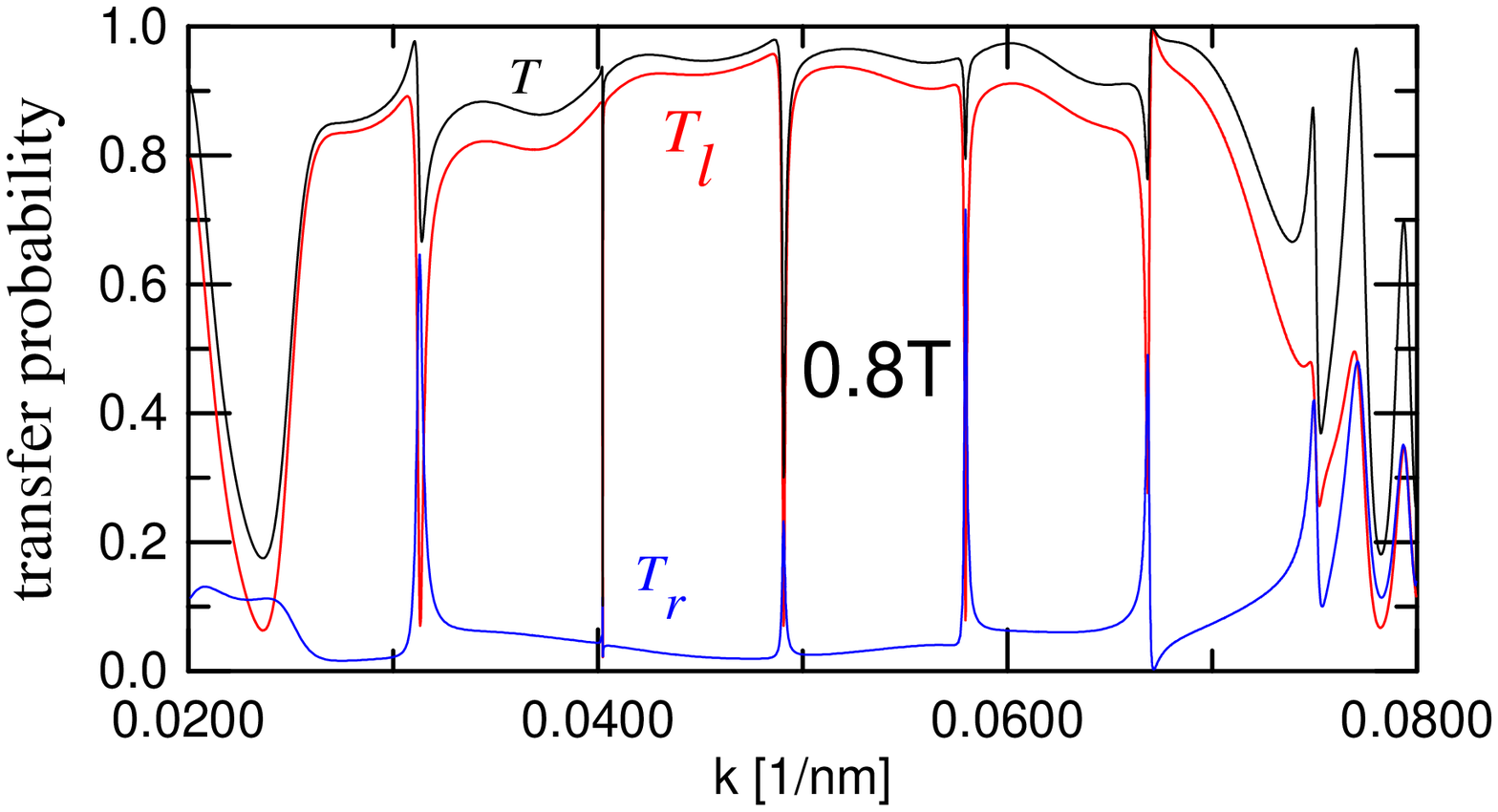}
    \end{tabular}
    \caption{
    Transfer probabilities to the left $T_l$ and right $T_r$ output channels and their sum $T$ as functions
    of the incident wave vector $k$ for $B=0$ (a), and $B=0.8$ T (b).
    }
    \label{wk}
    \end{figure}

    \begin{figure}[ht!]
     \hbox{\rotatebox{0}{\epsfxsize=70mm
                    \epsfbox[17 115 560 411] {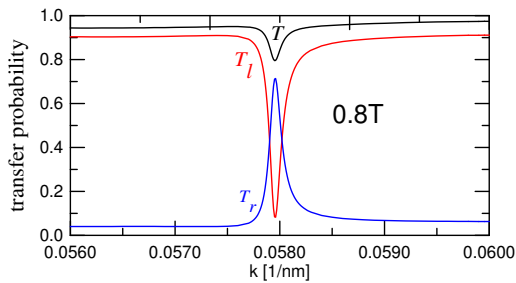}
                    \hfill}}
    \caption{Zoom of a fragment of Fig. \ref{wk}(b) }
    \label{zoom}
    \end{figure}

    \begin{figure*}[ht!]
     \hbox{\rotatebox{0}{\epsfxsize=190mm
                    \epsfbox[22 455 559 832] {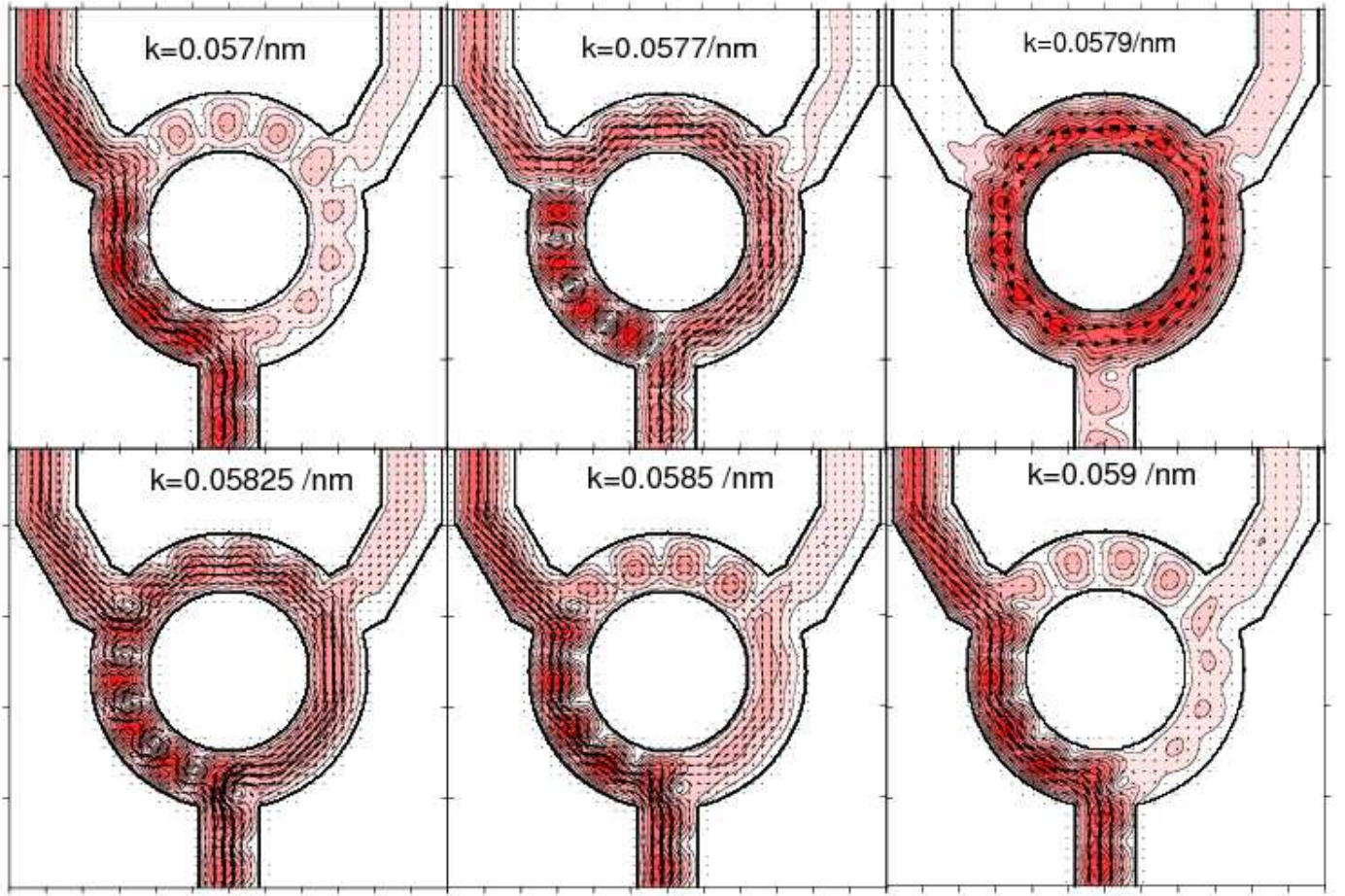}
                    \hfill}}
    \caption{The red contours show the absolute value of the wave function
    (the darker the shade of red - the larger $|\Psi|$) and probability current field (arrows) for $B=0.8$ T and several values of $k$
    indicated at the top of the figure.
    For the transfer probabilities see Fig. \ref{zoom}.
     }
    \label{ryj}
    \end{figure*}

    \subsection{Simulation of the temperature effects for broadening of the transport window}
    In order to estimate the effects of non-zero temperature for the transport
    we
    apply the linear response formula  for the conductance\cite{datta}
    \begin{equation}
    G=\frac{2e^2}{h} \overline{T},
    \end{equation}
    with
    \begin{equation}
    \overline{T}=\int T(E) \left(-\frac{\partial f}{\partial E}\right) dE,\label{tildet}
    \end{equation}
    and the Fermi function
    $f=\left(e^{(E-E_F)/k_B\tau}+1\right)^{-1}$,
    where $\tau$ stands here for the temperature.
    Formula ({\ref{tildet}) 
    accounts for averaging the transfer probability obtained in the Hamiltonian eigenstates within the transport window that is opened near the Fermi level by thermal excitations.  In the integral over the energy the wave vector $k$ corresponding to a given $E$ is found from the eigenequation
    (\ref{new2}).







    \begin{figure}[ht!]
    \begin{tabular}{ll}
     a) &  \epsfxsize=80mm \epsfbox[39 150 553 440] {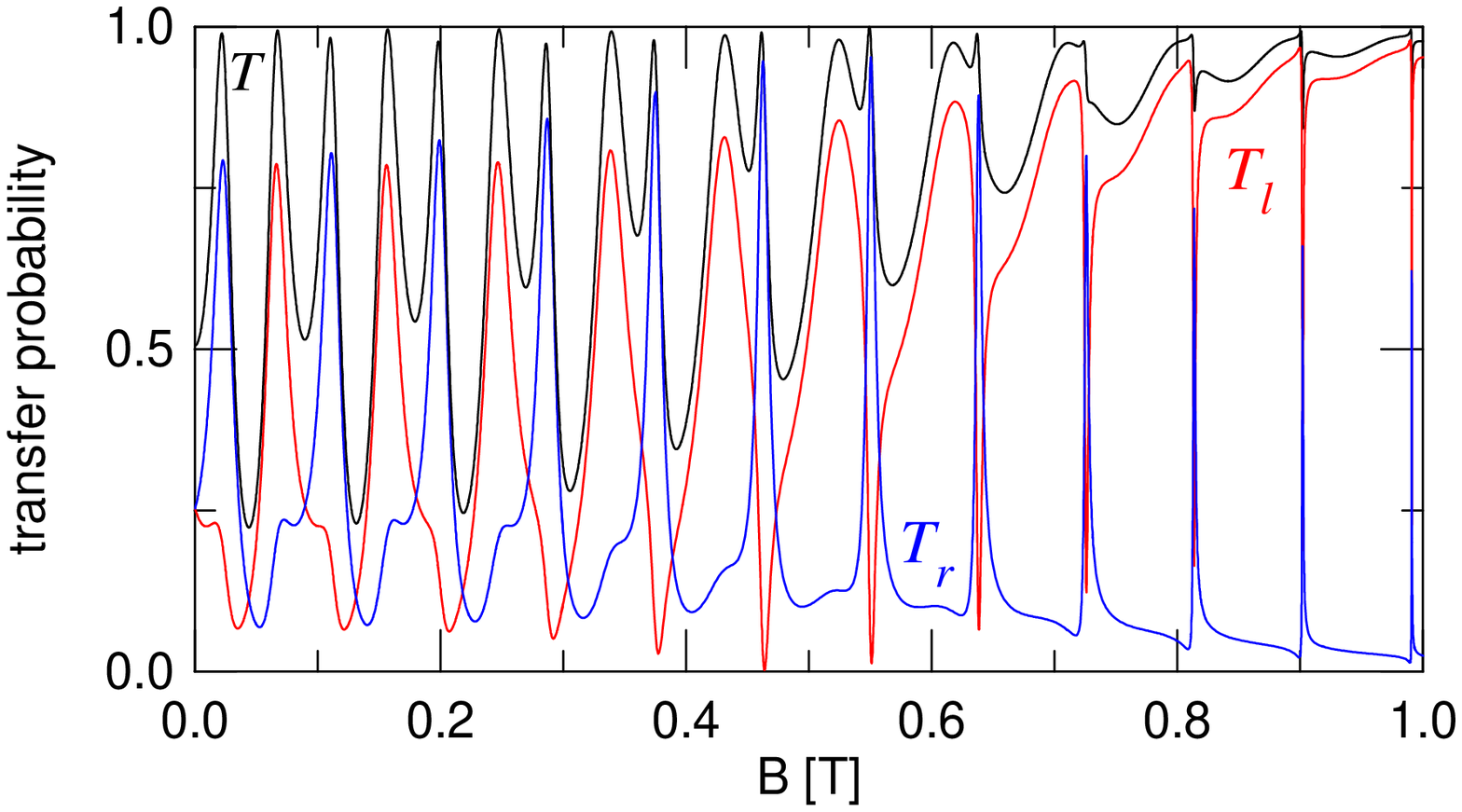} \\
     b) &   \epsfxsize=80mm  \epsfbox[39 150 553 440] {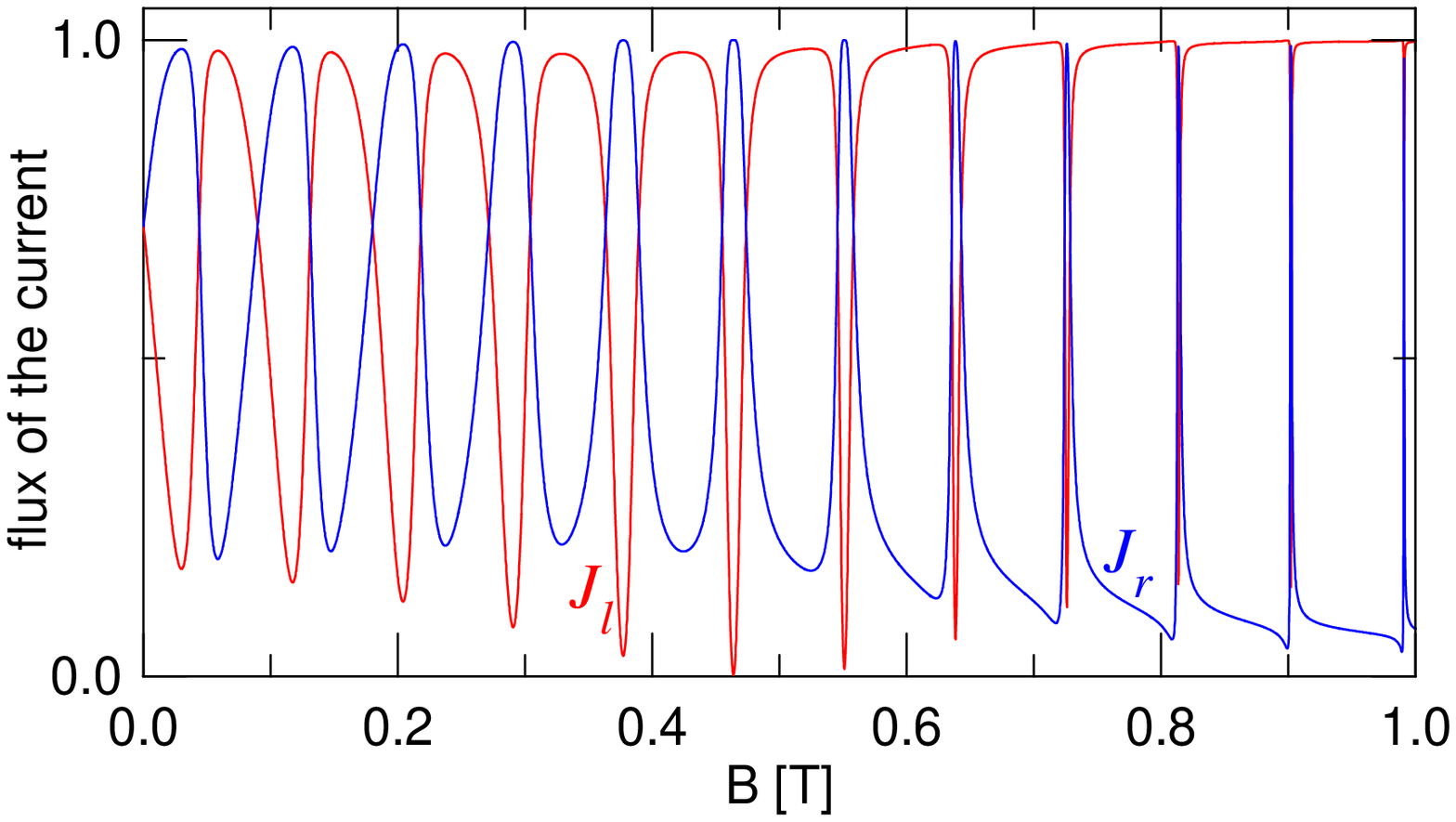}
    \end{tabular}
    \caption{
    (a) Transfer probabilities to the left $T_l$ and right $T_r$ output channels and their sum $T$ as functions
    of $B$ for $k=0.0683$ nm$^{-1}$.  (b) The flux of the current through the left and the right arms of the ring.
    }
    \label{wb683}
    \end{figure}

    \begin{figure}[ht!]
    \begin{tabular}{ll}
     a) &  \epsfxsize=80mm \epsfbox[39 150 553 440] {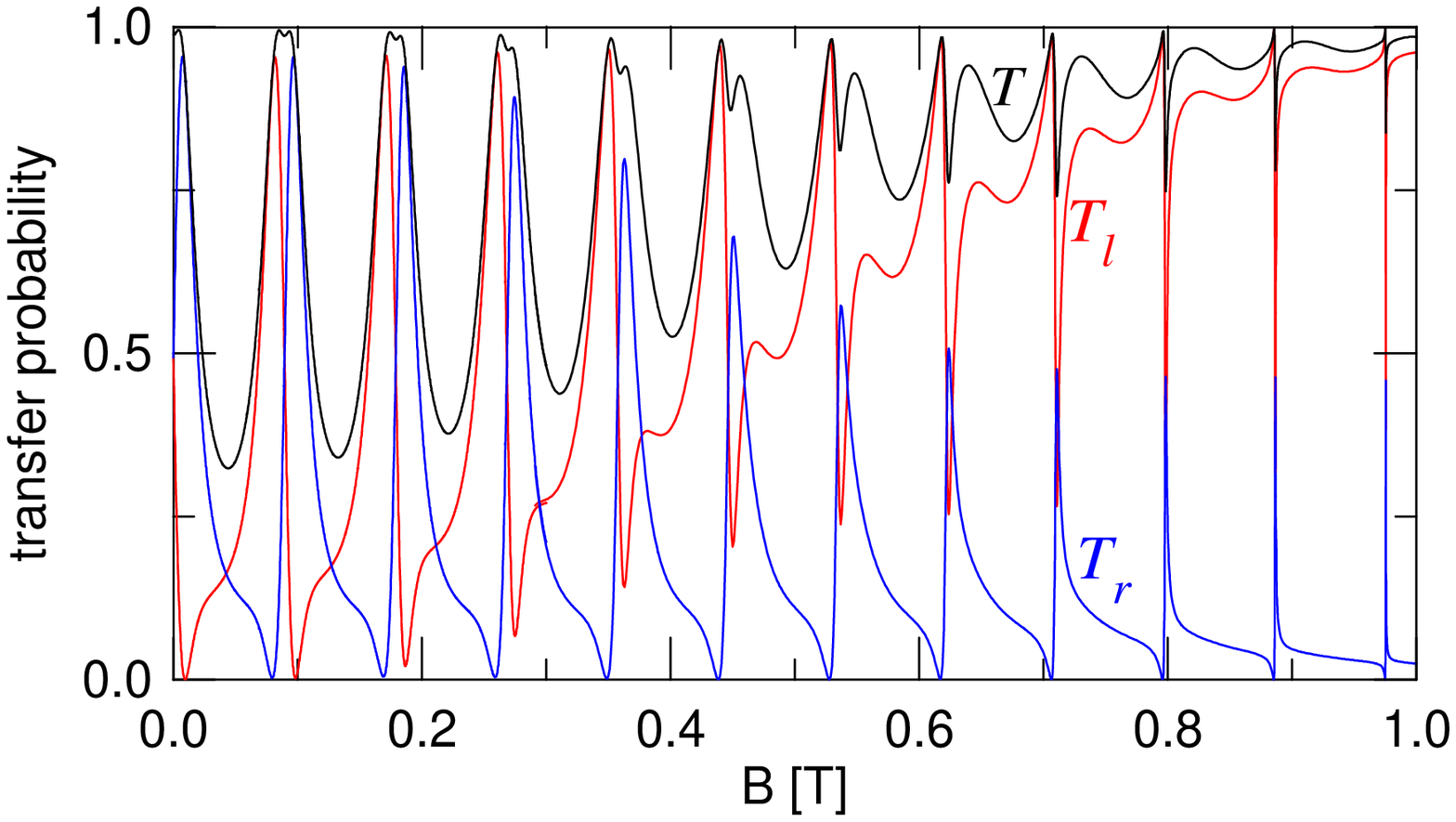} \\
     b) &   \epsfxsize=80mm  \epsfbox[39 150 553 440] {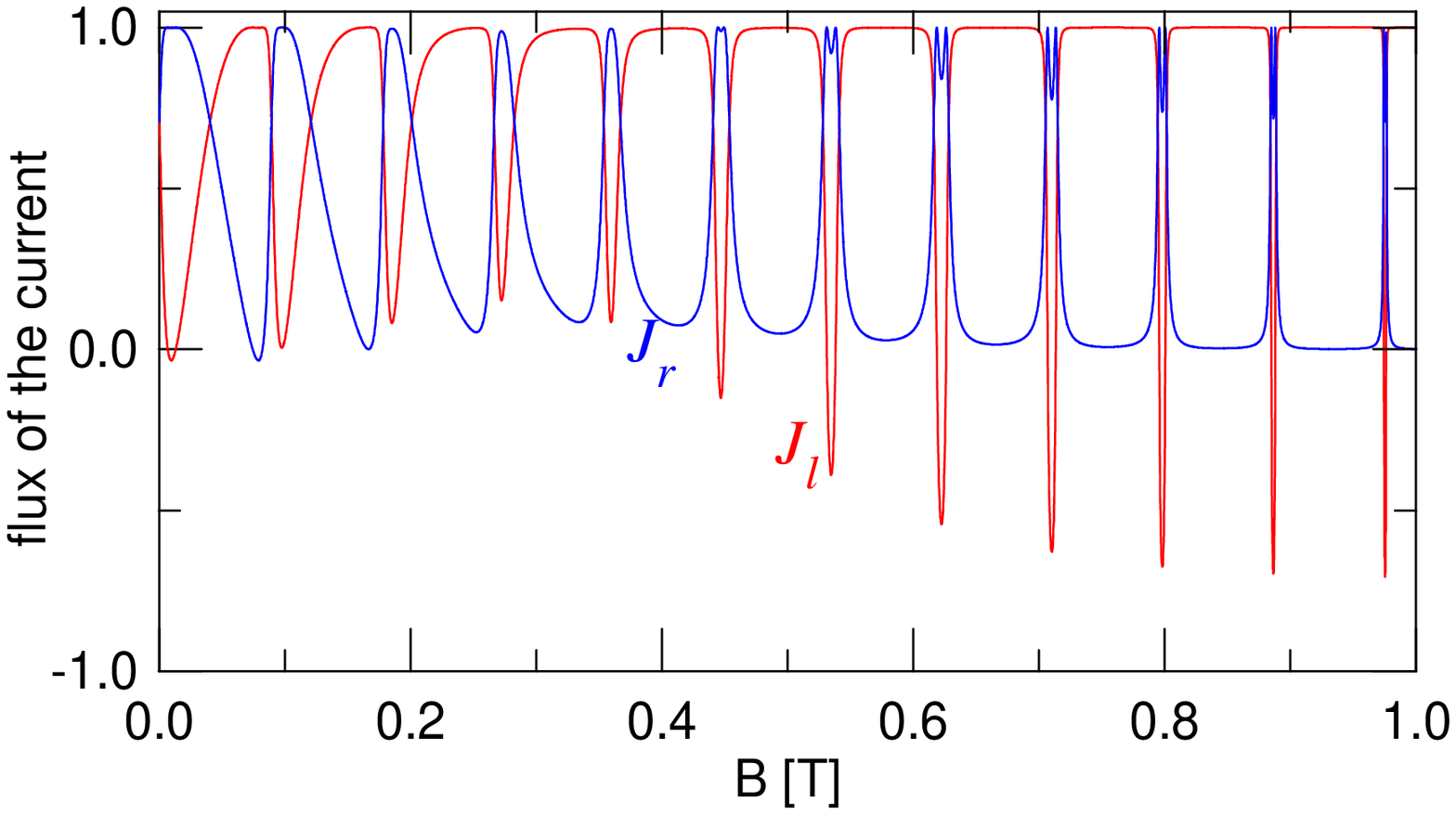}
    \end{tabular}
    \caption{
    Same as Fig. \ref{wb683} but for $k=0.0667$ nm$^{-1}$.
    }
    \label{wb667}
    \end{figure}

    \begin{figure}[ht!]
    \begin{tabular}{ll}
    a) \epsfxsize=67mm \epsfbox[39 150 553 440] {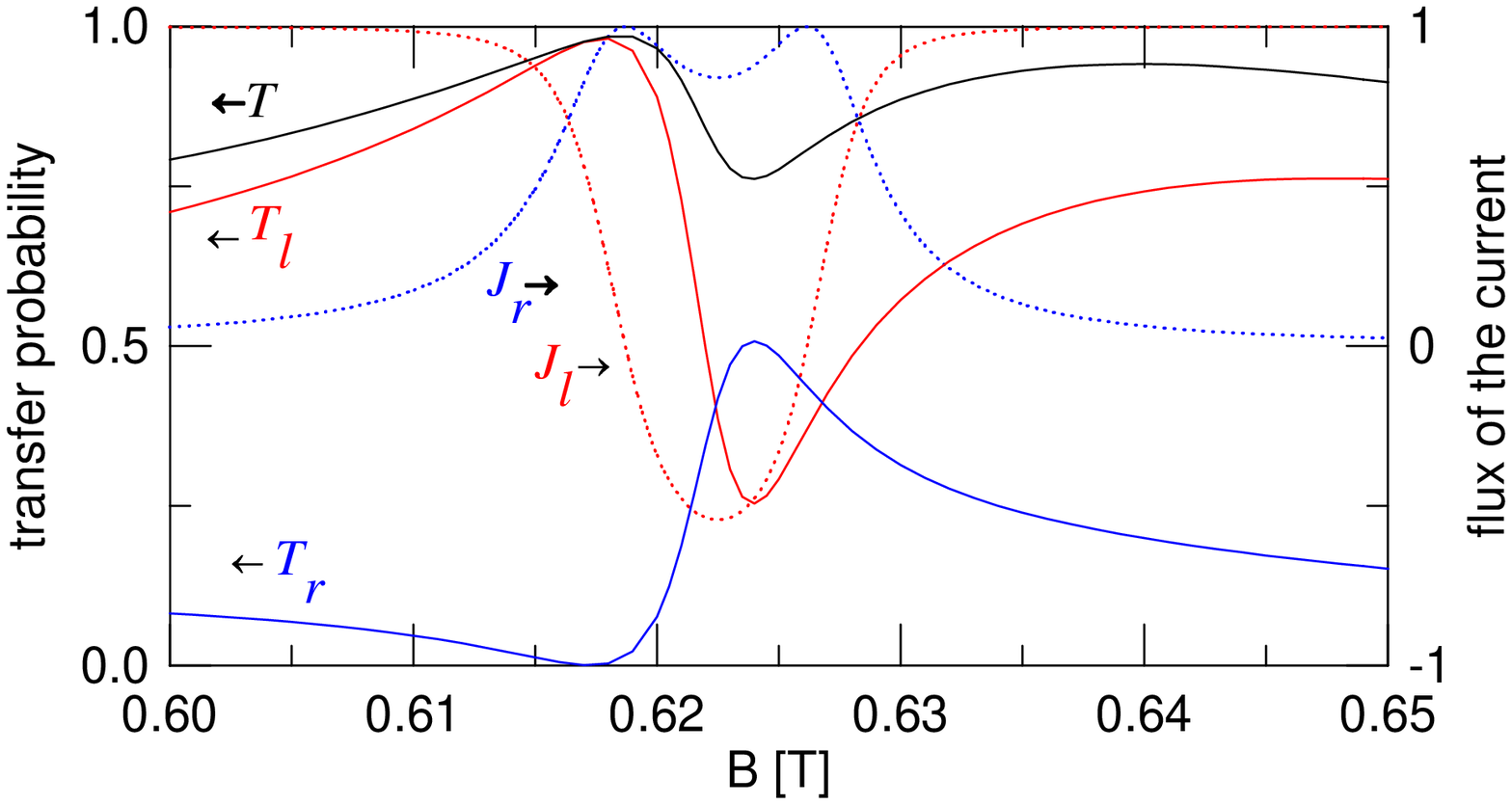} \\
    b) \epsfxsize=67mm \epsfbox[39 150 553 440] {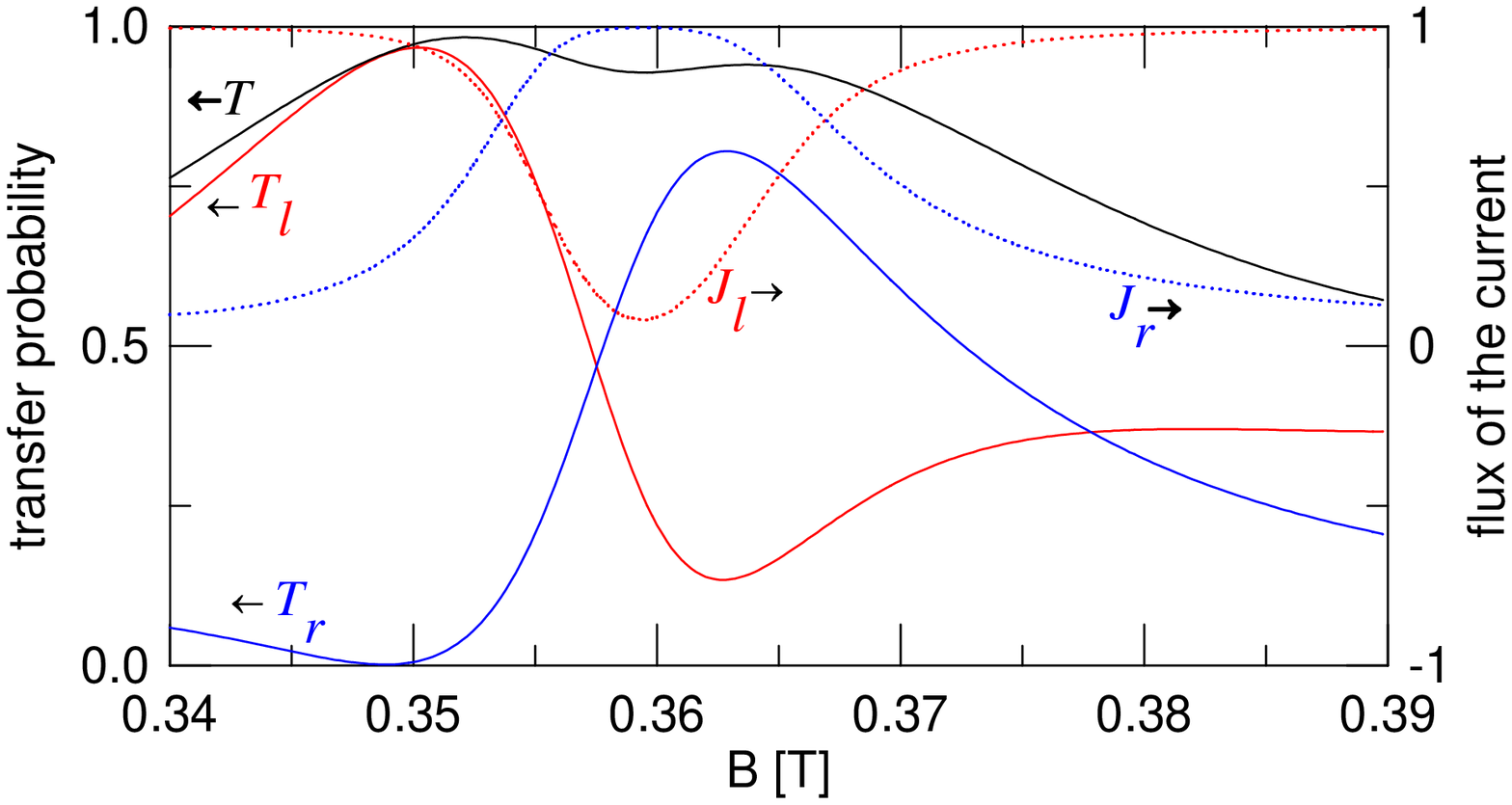}
     \end{tabular}
    \caption{
    Zoom of two fragments of Fig. \ref{wb667} corresponding to $T_r$ maxima. The solid lines show the transfer probabilities (left vertical axis),
    and the dotted ones the normalized flux of the probability density current through the left and right arms of the ring (right vertical axis).
    }
    \label{667f}
    \end{figure}

    \begin{figure*}[ht!]
     \hbox{\rotatebox{0}{\epsfxsize=180mm
                    \epsfbox[15 442 579 828] {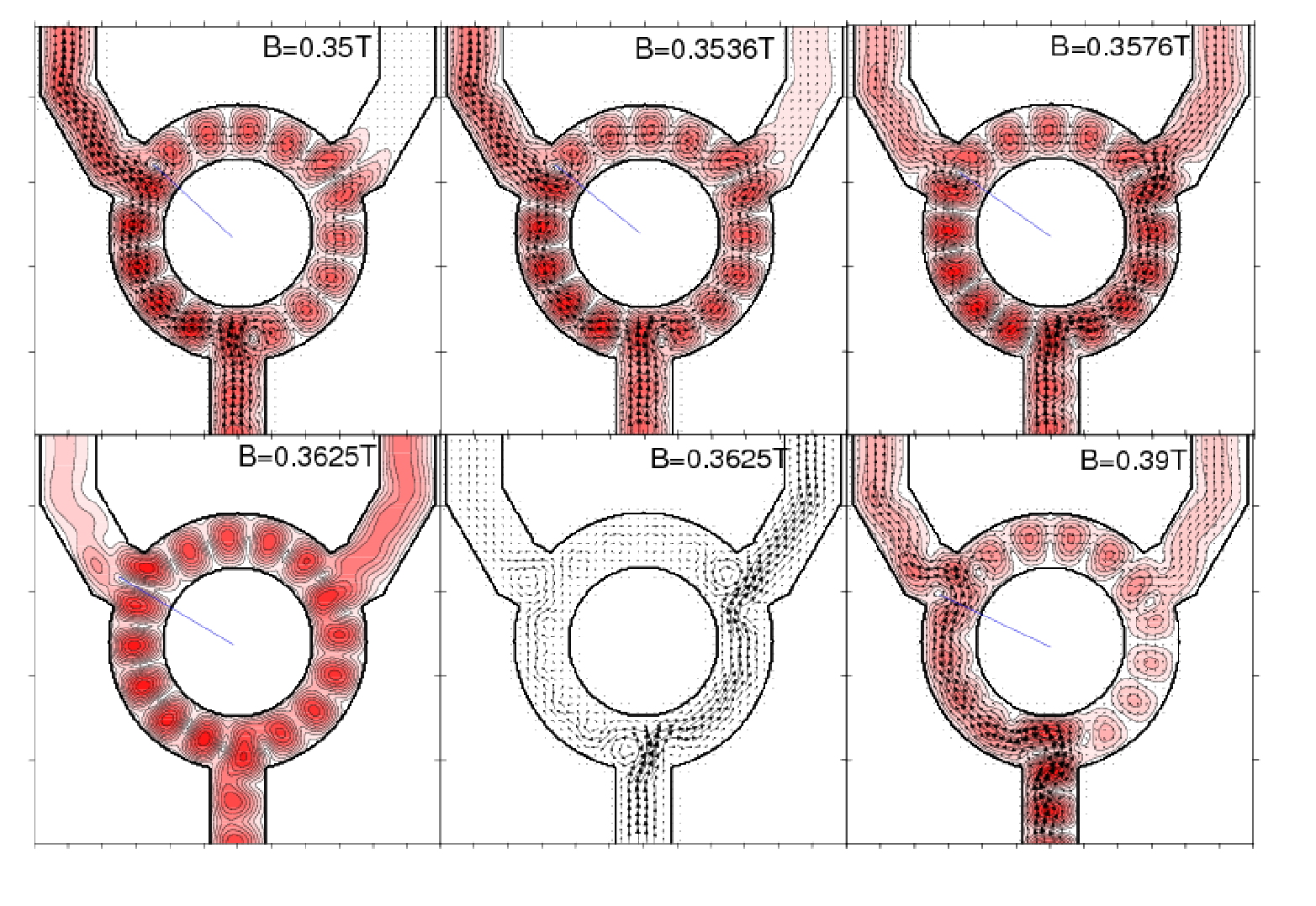}
                    \hfill}}
    \caption{The contour plot shows the absolute value of the wave function
    (the darker the shade of red - the larger $|\Psi|$) and probability current field (arrows) for  $k=0.0667$ /nm
    and values of the magnetic field of
     Fig. \ref{zoom}(b). }
    \label{ryj36}
    \end{figure*}

    \section{results and discussion}
    \subsection{results for 0 K}
    Calculated transfer probabilities as functions of the wave vector are presented in Fig. \ref{wk}.
    For $B=0$ one obtains $T_l(k)=T_r(k)$ due to the symmetry of the structure [Fig. \ref{wk}(a)].
    Non-zero magnetic field introduces asymmetry in the transfer probabilities [Fig. \ref{wk}(b)].
    Generally, at $B>0$ one observes that $T_l$ is enhanced at the expense of $T_r$ which is consistent with the orientation of the Lorentz force.
    Nevertheless, for discrete values of $k$ sharp dips
    of $T$ appear at higher $B$ [Fig. \ref{wk}(b)]. The dips of $T$ coincide with the minima of  $T_l$ and peaks of $T_r$.
    A zoom of one of $T$ dips is shown in Fig. \ref{zoom}.
    The amplitude of the wave function and probability current distribution for $k$ near the dip are displayed in Fig. \ref{ryj}.
    For $k=0.057$ nm$^{-1}$ the electron is directed to the left arm of the ring and then to the left output channel as previously
    described by the time dependent calculations.\cite{time,epl,poniedzialek} For  $k=0.0577$ nm$^{-1}$ the current forms vortices in the left arm and  the actual electron transfer occurs through the right arm of the ring. For  $k=0.0577$ nm$^{-1}$ the current goes through the right arm but the electron transfer to the right
     lead has still low probability.
    For $k=0.0579$ nm$^{-1}$ -- at the center of the $T_l$ dip ($T_r$ peak) -- the current forms a giant counterclockwise vortex around the entire ring.
    A minimum of the wave function amplitude is formed at the center of the entrance to the left output channel - similar to the one observed for $k=0.0577$ nm$^{-1}$
    at the right output channel.
    For larger $k$  the current restarts to flow through the left arm as guided by the classical magnetic forces.

    The magnetic forces influence the distribution of the charge density within the ring.
    In Fig. \ref{ryj} we observe a distinct shifts of the wave function amplitude with respect to the axes of the channels
    correlated with the direction of the current and consistent with the orientation of the Lorentz force. For $k=0.057$ nm$^{-1}$ the wave function is distinctly shifted
    to the left edge of the input and the output channels as well as to the external edge of the left arm of the ring.
    For $k=0.0577$ nm$^{-1}$, when the transfer of the current through the left arm is blocked, the wave function maxima
    between the input and left output lead  are placed symmetrically between the internal and external edges of the ring.
    For the giant anticlockwise vortex found for $k=0.0579$ nm$^{-1}$ the wave function is pushed to the inner edge of the ring.

    %

    In experiments the conductance is usually measured in function of the magnetic field.
    Fig. \ref{wb683}(a) shows the transfer probabilities as functions of $B$ for
     $k=0.0683$ nm$^{-1}$.
    At low $B$ the maxima of $T$ correspond to interlaced
    peaks of $T_l$ and  $T_r$.
    At higher $B$ the value of $T_l$ increases on average and the peaks of $T_r$ become very narrow.
    Pronounced dips of $T_l$ are formed at the positions of $T_r$ maxima. The peak/dip structure
    occurs periodically with the spacings of $\Delta B=0.09$ T which corresponds to the flux quantum
    threading the one-dimensional ring of an effective radius 121 nm that well agrees with the geometry of
    the model structure (Fig. 1).

    In order to quantify the direction of the current flow within the ring
    we calculate the flux of the current at the horizontal cross section of the arms of the ring  $y=240$ nm (see Fig. 1). The fluxes are then normalized to obtain $J_l^2+J_r^2=1$.
    In Fig. \ref{wb683}(b) we notice that for larger $B$ outside $T$ dips nearly all the current goes
    through the left arm of the ring. 

    Fig. \ref{wb667} corresponds to $k=0.0667$ nm$^{-1}$ for which a maximum of $T=2T_l=2T_r$ is found for $B=0$ [see Fig. \ref{wk}(a)].
    At low $B$ the peaks of $T_l$ and $T_r$ appear very close to one another forming a wider $T$ maxima.
    For higher $B$ i) the maxima of $T_r$ turn into narrow peaks which coincide with the dips of $T_l$
    ii) outside the $T$ dips the current flows up through the left arm of the ring while the current flux through the right arm is close to zero,
    as discussed above for $k=0.0683$ nm$^{-1}$.

    Enlarged fragments of Fig. \ref{wb667} corresponding to two dips of $T_l$ are shown in Fig. \ref{667f}. The amplitude of the wave function
    and the probability density current for the $B$ range of Fig. \ref{667f}(b) is illustrated in Fig. \ref{ryj36}.
    For $B=0.35$ T the transfer probability to the left lead is maximal, while $T_r$ is minimal. The current goes nearly entirely by the left arm.
    Note the pronounced elongated minimum of the wave function at the exit to the right output channel.
    This wave function would be effectively coupled to the second subband of the right channel, but
    the latter corresponds to a much higher energy, so the transfer to the right lead is blocked.
    For $B=0.3536$ T a leakage of the current to the right lead is observed and $J_r$ becomes equal to $J_l$. For $B=0.3576$ T
    the vortices of the current appear in the left arm. The transfer of the current through the left arm
    is nearly blocked.
    Note the position of the sharp minimum of the wave function near the left output lead pointed by the blue line in Fig. \ref{ryj36}.
    As $B$ grows from $0.35$ T the minimum is shifted to the left and for $B=0.3625$ T it is found at the center of the
    junction of the left output channel to the ring. For this value of the magnetic field $T_l$ becomes minimal.
    Generally in our simulations a minimum of the wave function amplitude at the center of the junction
    to the left output lead is found for all minima of $T_l$ which become sharp at higher $B$.
    When the electron transfer to the left lead is blocked or hampered, the current goes to
    the right output channel leading to appearance of a maximum of $T_r$.
    For $B=0.3625$ T the current forms vortices in the left arm as well as between the output leads and the main electron transfer goes through the
    right arm to the right output lead. For $B=0.39$ T the minimum of the wave function is shifted to the lower edge of the left junction and the
    current transfer through the left arm to the left output channel restarts.


    For $J_r>J_l$ the direction of the current circulation is opposite to the one preferred by the Lorentz force.
    Intervals of $B$ corresponding to this orientation of the current become narrow at higher field [see Figs. \ref{wb683}(b) and \ref{wb667}(b)].
    Also the magnetic field interval for which $T_r>T_l$ become narrower at higher field [see also Fig. \ref{667f}].

    In order to conclude this section we note, that at higher magnetic field  the electron transfer goes predominantly
    through the left arm of the ring to the left output lead, as should be expected due to the orientation of the Lorentz force.
    For narrow intervals of $k$ or $B$ wave function interference within the ring
    leads to formation of a wave function minimum at the entrance to the left output channel which
    blocks the transfer to the left lead. The $T_l$ minima are associated
    with reversal of the current circulation and appearance of $T_r$ maxima which turn into sharp peaks at higher $B$.

    \begin{figure}[ht!]
    \begin{tabular}{ll}
     a) &  \epsfxsize=60mm \epsfbox[31 164 526 483] {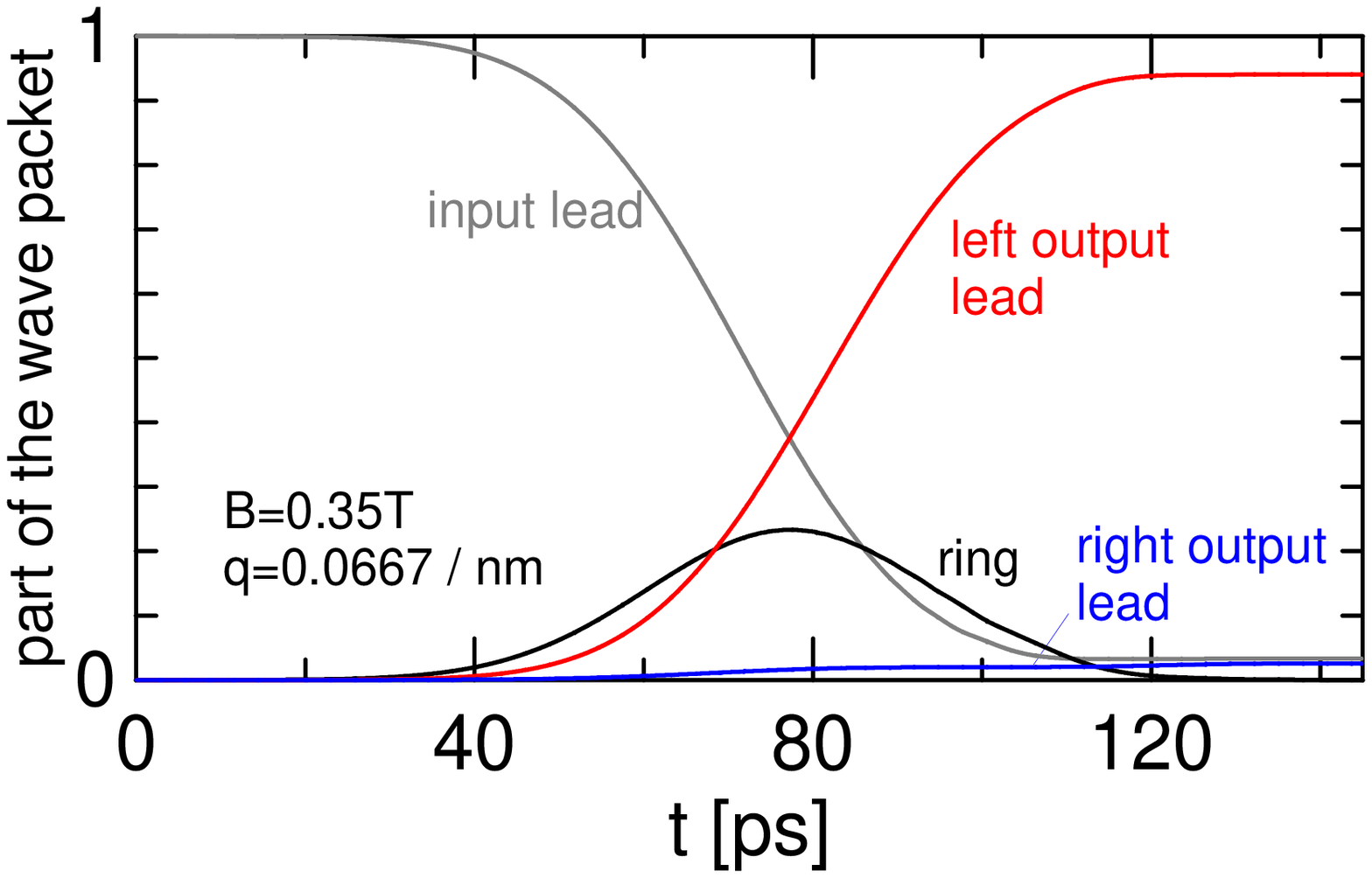} \\
     b) &   \epsfxsize=60mm  \epsfbox[31 164 526 483] {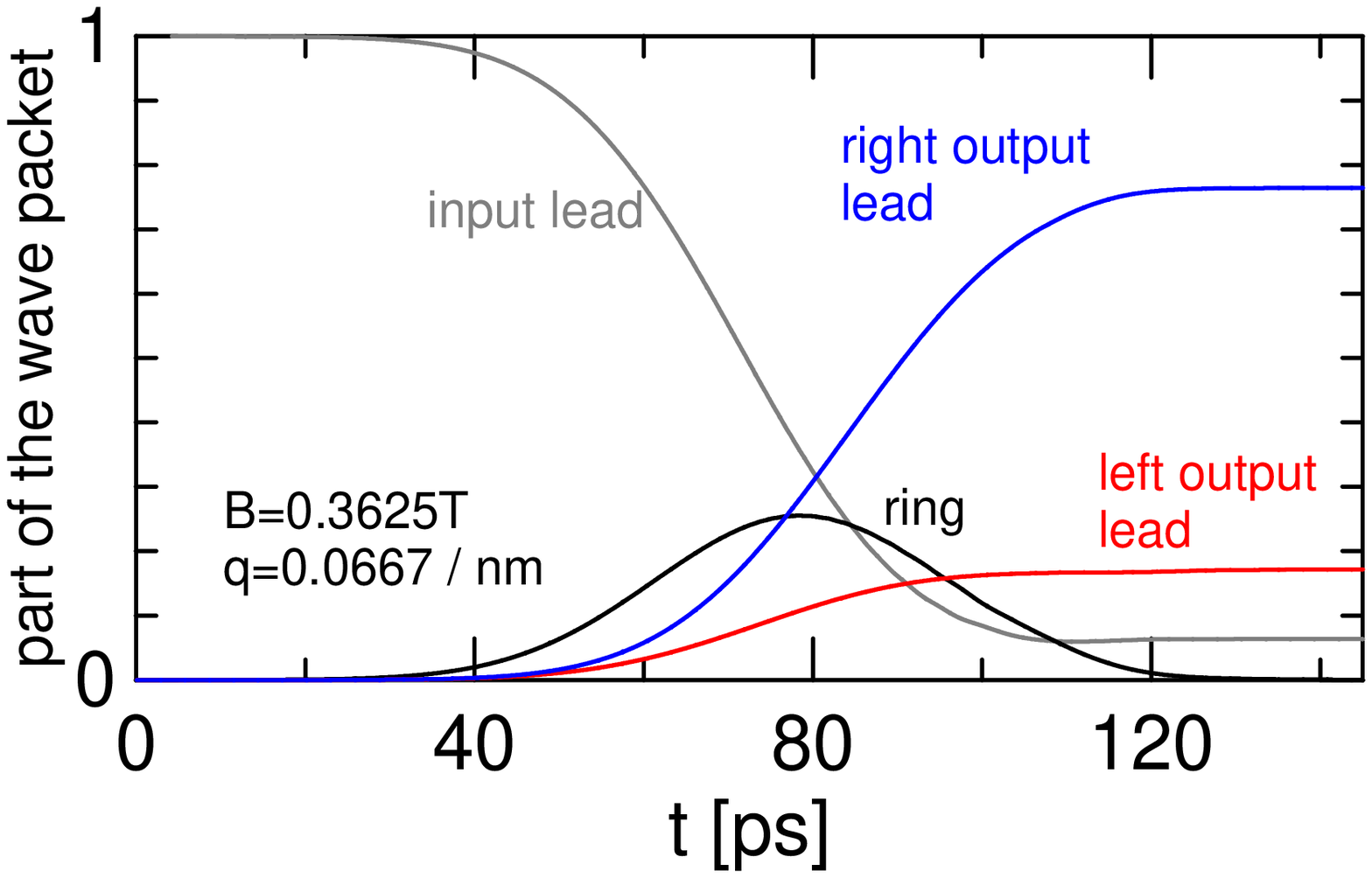} \\
     c) & \epsfxsize=60mm  \epsfbox[31 164 526 483] {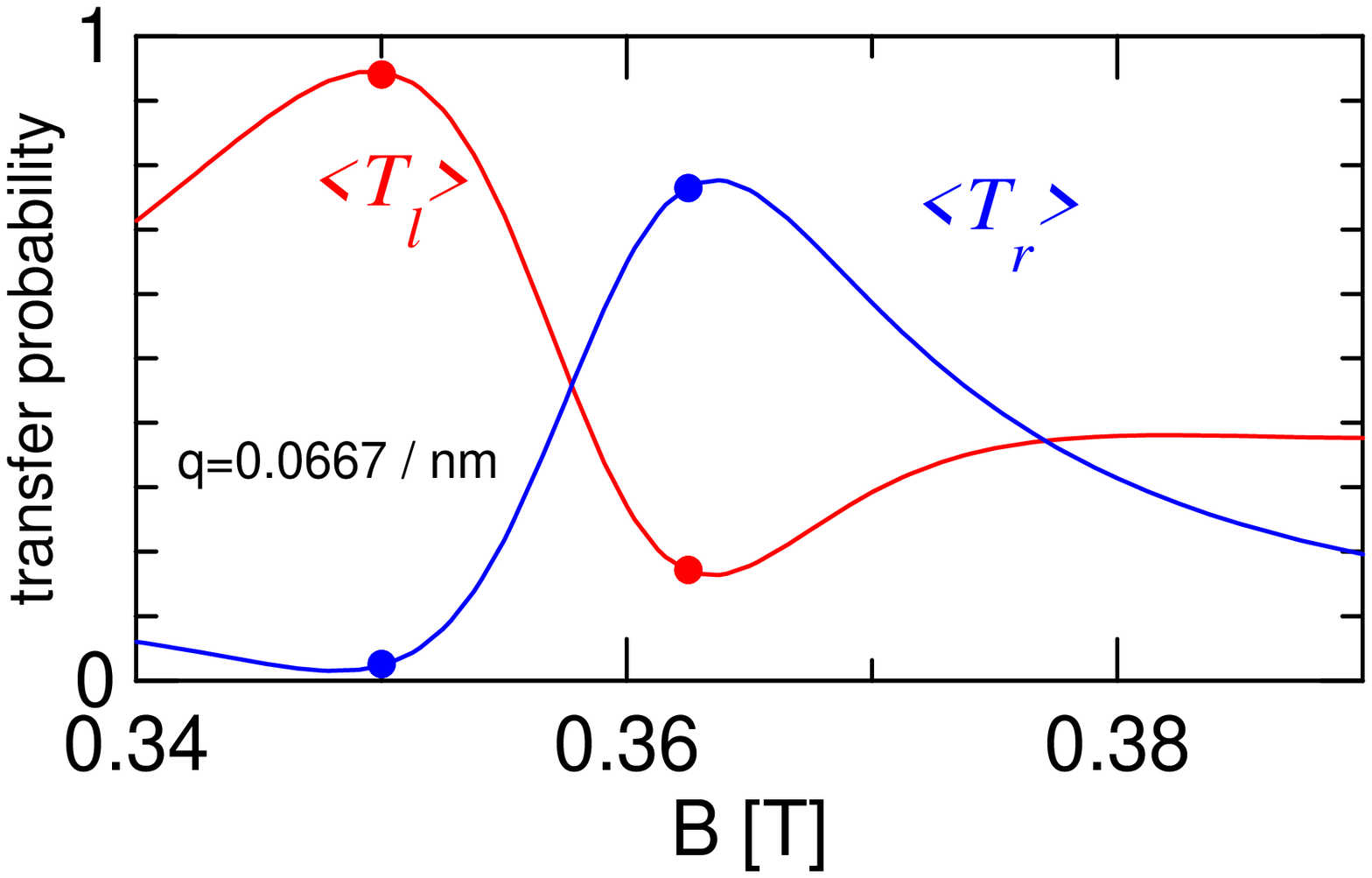}
    \end{tabular}
    \caption{Parts of the wave packet in the input and the output leads as well as within the ring for $q=0.0667$ nm$^{-1}$ in the magnetic field of
    $B=0.35$ T (a) and 0.3625 T (b).
    In (c) the lines show the transfer probabilities as functions of the magnetic field
    obtained by a time-independent calculation in which the transfer probabilities are averaged over the Gaussian distribution
    corresponding to the wave packet [Eq. (\ref{lol})]. The dots show the results of the time dependent calculation: the parts of the wave packet that are found in the output leads at the end
    of simulation for both $B$ considered in (a) and (b).
    }
    \label{timepart}
    \end{figure}

    \begin{figure*}[ht!]
     \hbox{\rotatebox{0}{\epsfxsize=140mm
                    \epsfbox[111 224 705 707] {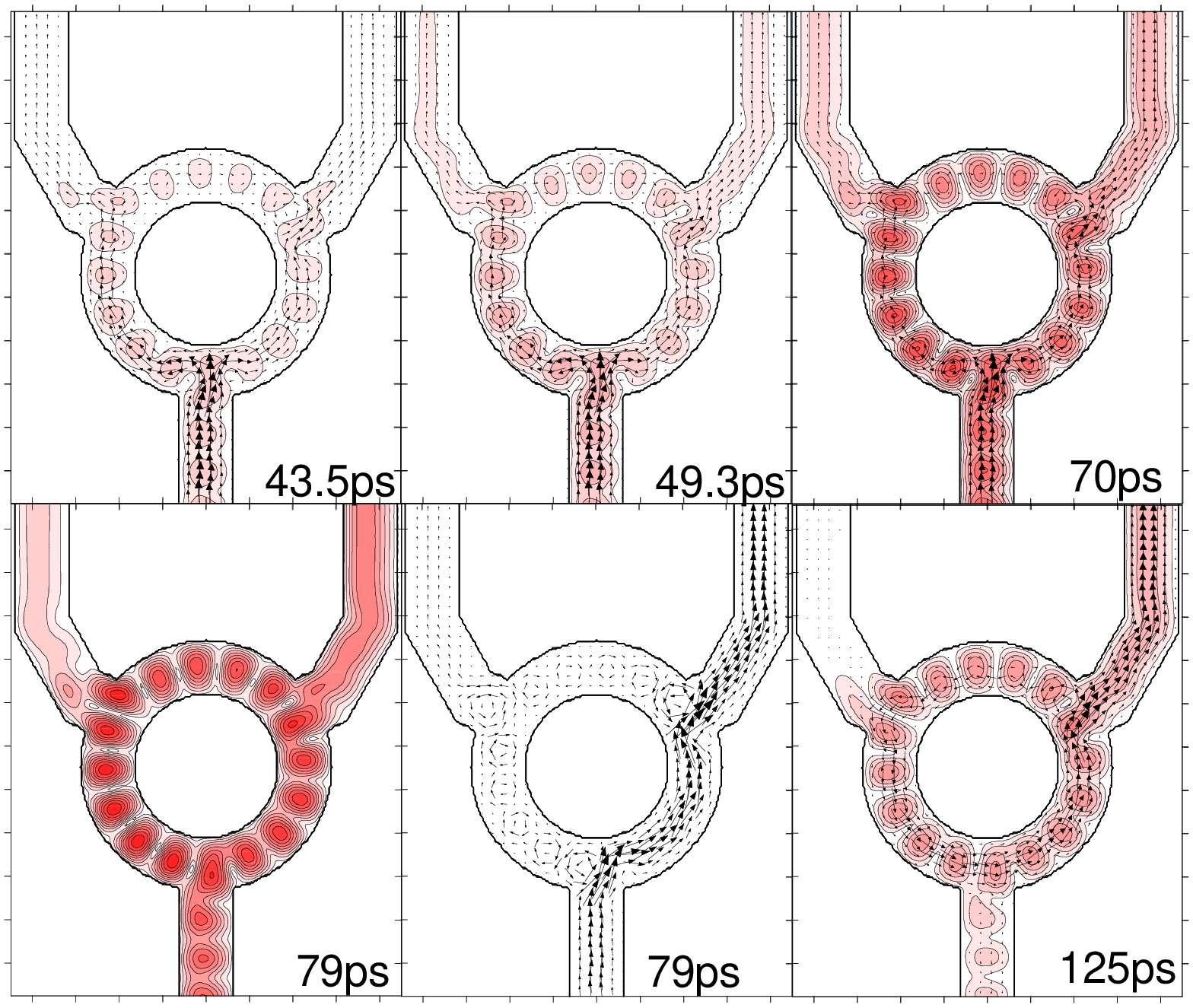}
                    \hfill}}
    \caption{Snapshots of the time-dependent simulation for the average wave vector $q=0.0667$ nm$^{-1}$
    and $\Delta k=5.5\times 10^{-4}$ nm$^{-1}$ [see Eq. (\ref{ic})] for chosen moments in time.
     The contour plots show the amplitude of the wave
    function and the arrows -- the current distribution. The color scale for the amplitude is the same for all
    the plots. The scale for the current vectors is different in each plot. }
    \label{td}
    \end{figure*}

    \subsection{Wave packet simulation}
    The results presented so far indicate that for some intervals of the magnetic field the current flows in the opposite direction to the one
    indicated by the Lorentz force.
    The results of the wave packet simulation for nearly definite values of the packet wave vector should provide the transfer probabilities close to the ones
    found for  the Hamiltonian eigenstates.
    However, by the Ehrenfest theorem in the wave packet dynamics the average values of electron momentum and position
    follow classical laws. Hence, for $B>0$ a preferential injection of the packet into the left arm of the ring is should be expected for any magnetic field,
    on the contrary to the anomalous current injection that is found for Hamiltonian eigenstates for some values of $B$. In order to inspect this contradiction
      closer we performed wave packet simulations, in which we
    assume $\Delta k=5.5\times 10^{-4}$ nm [see Eq. (\ref{ic})]. 
    This wave vector dispersion for the studied structure and $k_F=0.0667$ nm$^{-1}$
    corresponds roughly to the thermal widening of the transport window which occurs at $150$ mK.
    The spatial spread of the initial wave function is then as large as 4 $\mu$m and we localize the wave packet  $Y=-8$ $\mu$m below
    the ring in the initial condition [Eq.(\ref{ic})].

    Figure {\ref{timepart}(a,b) shows the parts of the wave packet in the leads and within the ring for $B=0.35$ T and $B=0.3625$ T.
    In Fig. \ref{timepart}(b) we notice an enhanced packet transfer
    to the right output lead in consistence with Fig. \ref{667f}(b).
    Fig. \ref{td} shows the snapshots of the wave function amplitude
    and the probability current distributions for $B=0.3625$ T. When wave packet enters the ring
    more of the electron wave function goes into the left lead ($t=43.5$ ps and $t=49.3$ ps).  At $t=70$ ps
    an elongated wave function minimum is found at the entrance to the left lead. The current flow to the right output lead is visibly enhanced.
    For $t=79$ ps the part of the wave packet inside the ring is maximal and we find that both the wave function amplitude and
    the current distributions are very close to those found in the Hamiltonian eigenstate for $k=0.0667$ nm [see Fig. \ref{ryj36} for $B=0.3525$ T].

    Summarizing, in the time dependent simulations with a nearly monoenergetic wave packet one first observes an  asymmetric injection of the packet to the arms of the ring in accordance
    with the Lorentz force orientation. Next the interference conditions similar to the ones found in the Hamiltonian eigenstates
    are formed. For  $B=0.3525$ T the interference blocks the electron transfer to the left lead.

    The presented results of the wave packet simulation were obtained for an extremely low value of $\Delta k$.
    The time dependent simulations
    are useful for observation of
    the enhanced electron transfer to the right lead only for relatively low values of $B$, before the $T_r$ maxima turn into peaks
    as sharp as in Fig. \ref{wb667}(a) for $B=0.8$ T.
    The  $\Delta k$ applied here corresponds to roughly 1/7 of the length of horizontal axis of Fig. \ref{zoom},
    which largely exceeds the width of the $T_l$ dip.

    In Fig. \ref{timepart}(c) we compared the transfer probabilities
    estimated by the wave packet simulation with the ones obtained by the time-independent approach after
    calculating an averaged over the wave packet probability density in $k$ space, i.e.,
    \begin{equation}
    \langle T\rangle=C\int dk T(k)\exp\left(-2(k-q)^2/\Delta k^2\right).  \label{lol}
    \end{equation}
    Fig. \ref{timepart}(c) shows that
    the results of the wave packet simulations are consistent with the $k$-vector averaged transfer probability as calculated
    for Hamiltonian eigenstates.

    \begin{figure*}[ht!]
    \begin{tabular}{llll}
     a) &  \epsfxsize=70mm \epsfbox[31 98 555 391] {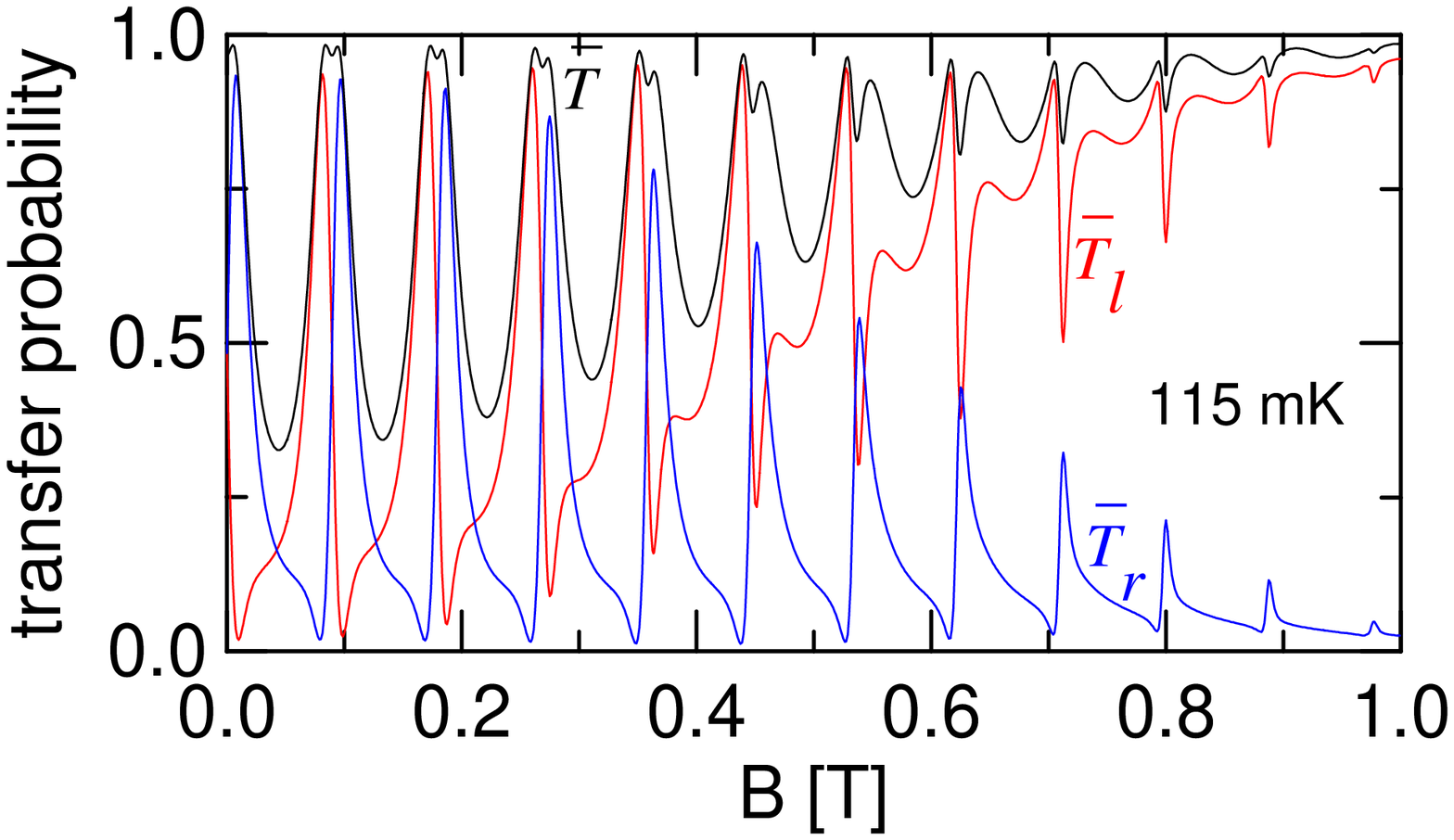} & \epsfxsize=70mm \epsfbox[31 98 555 391] {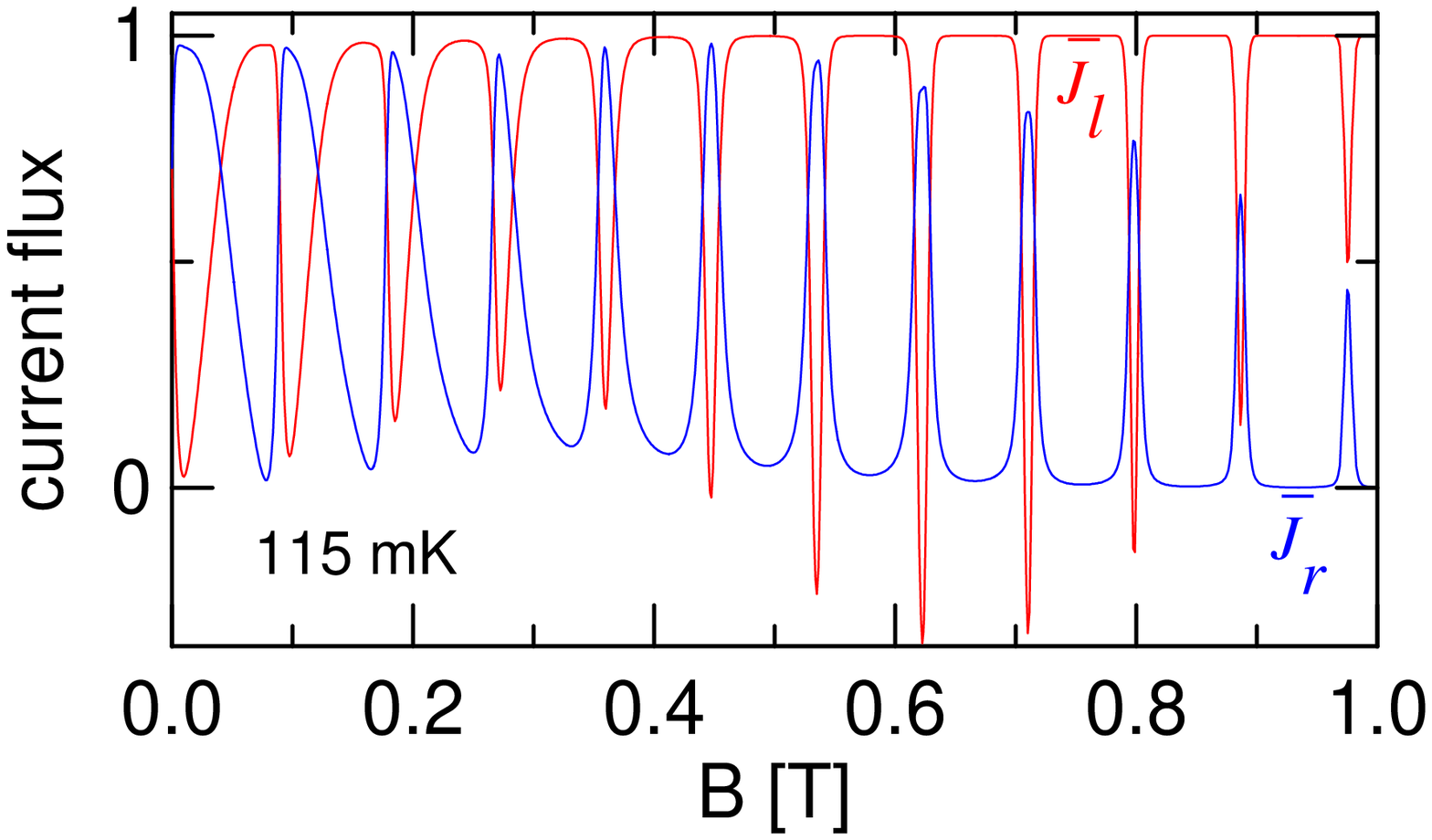}& b) \\
     c) &  \epsfxsize=70mm \epsfbox[31 98 555 391] {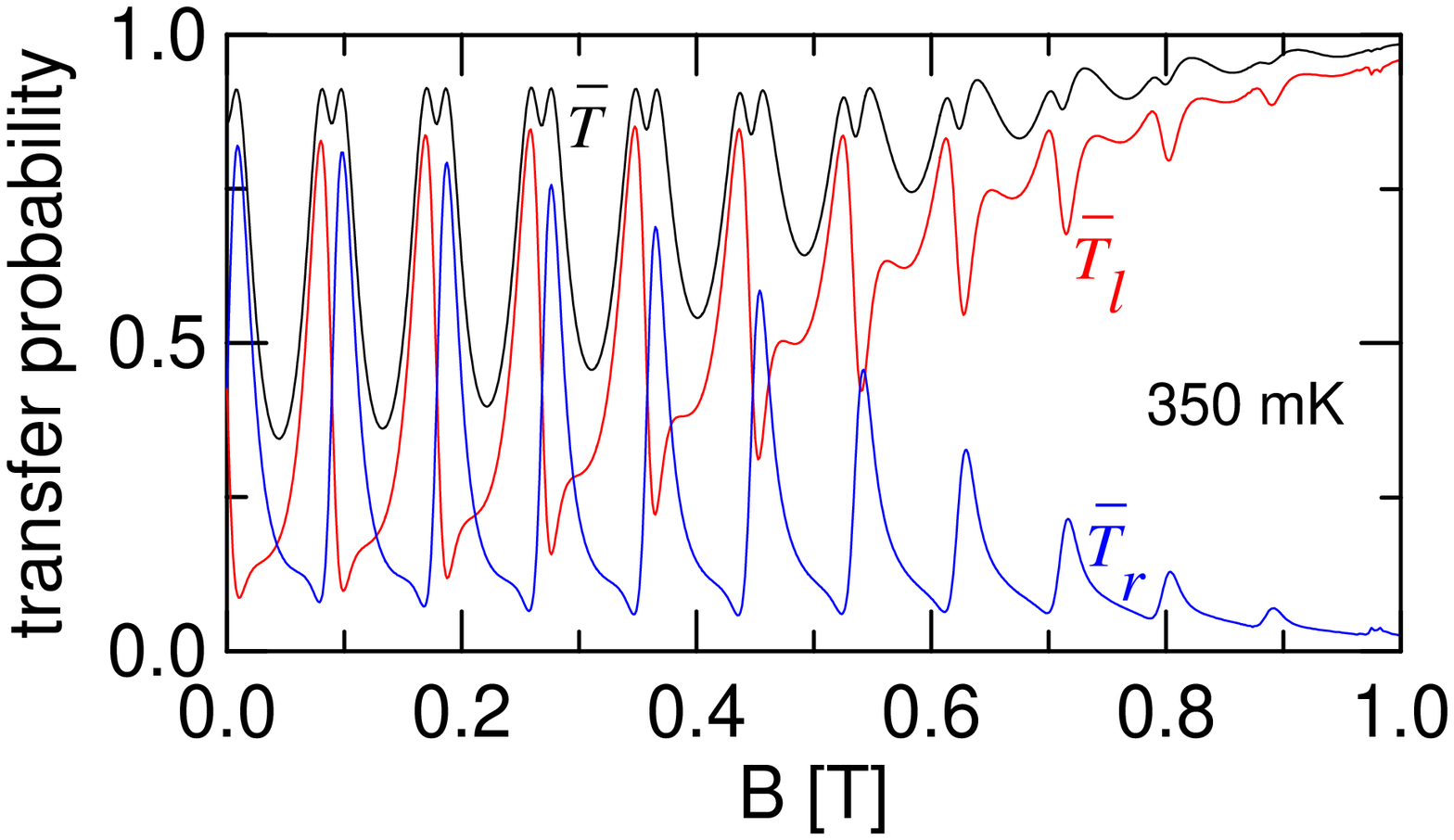} & \epsfxsize=70mm \epsfbox[31 98 555 391] {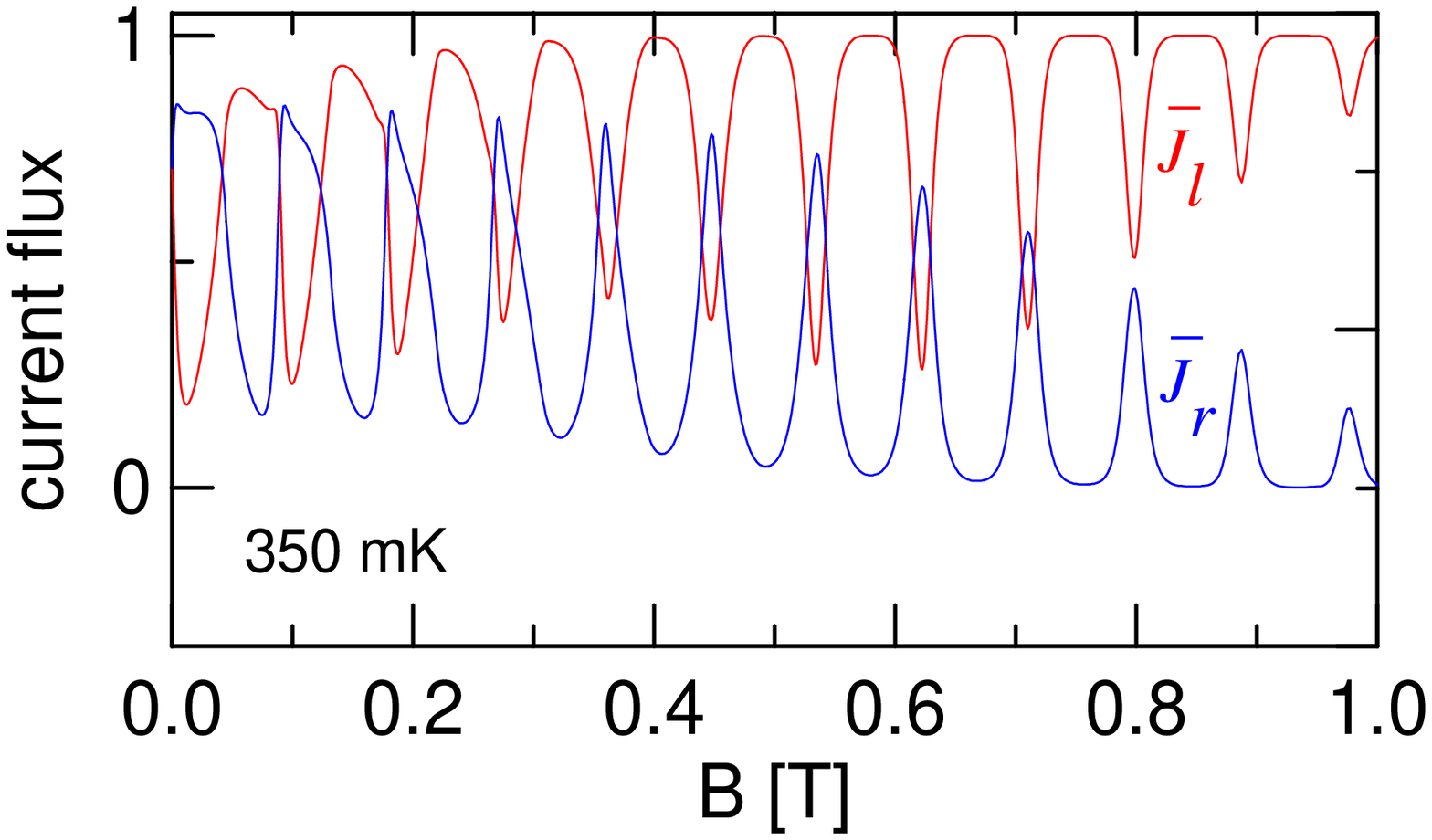}& d) \\
      e) &  \epsfxsize=70mm \epsfbox[31 98 555 391] {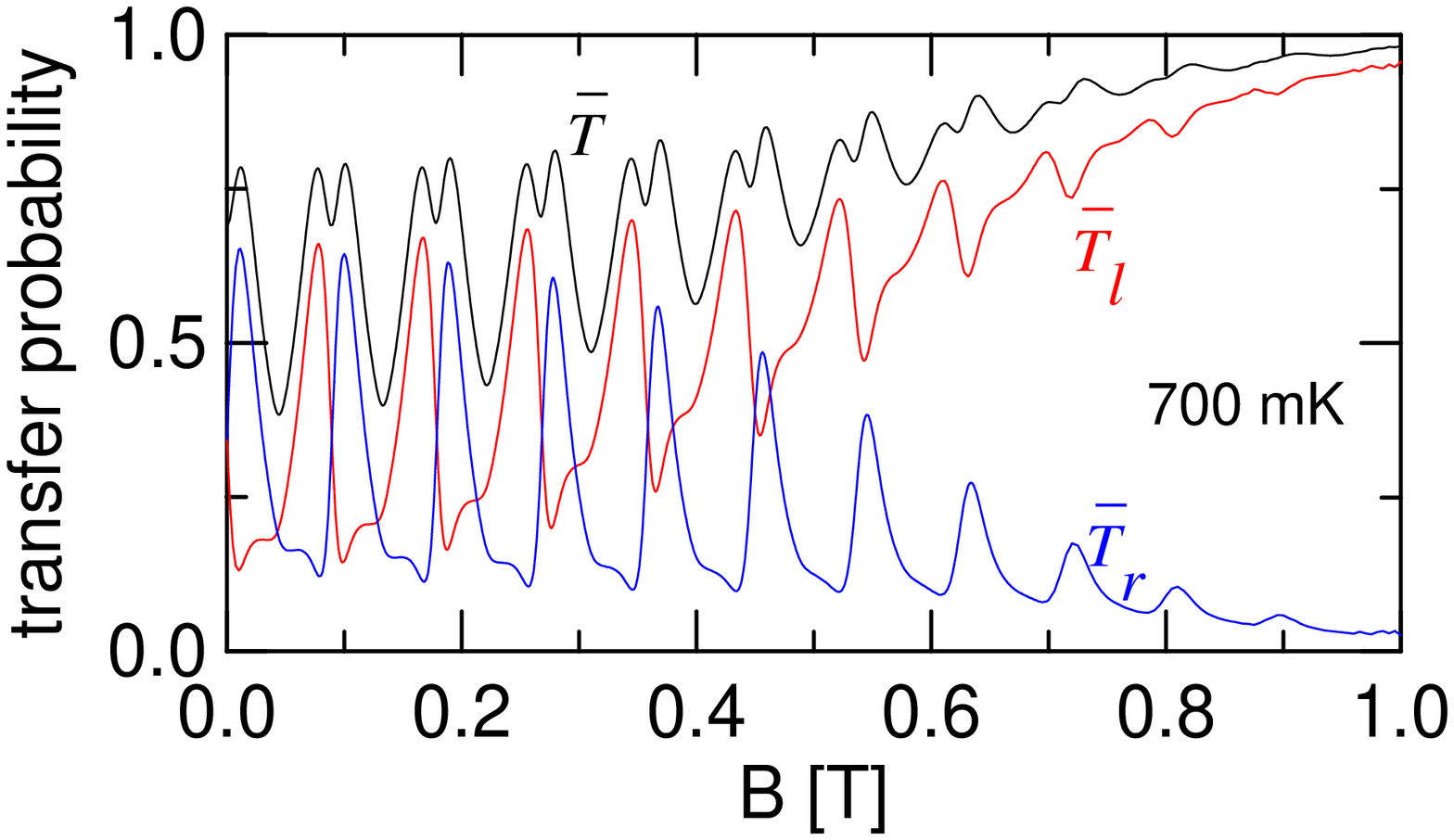}  & \epsfxsize=70mm \epsfbox[31 98 555 391] {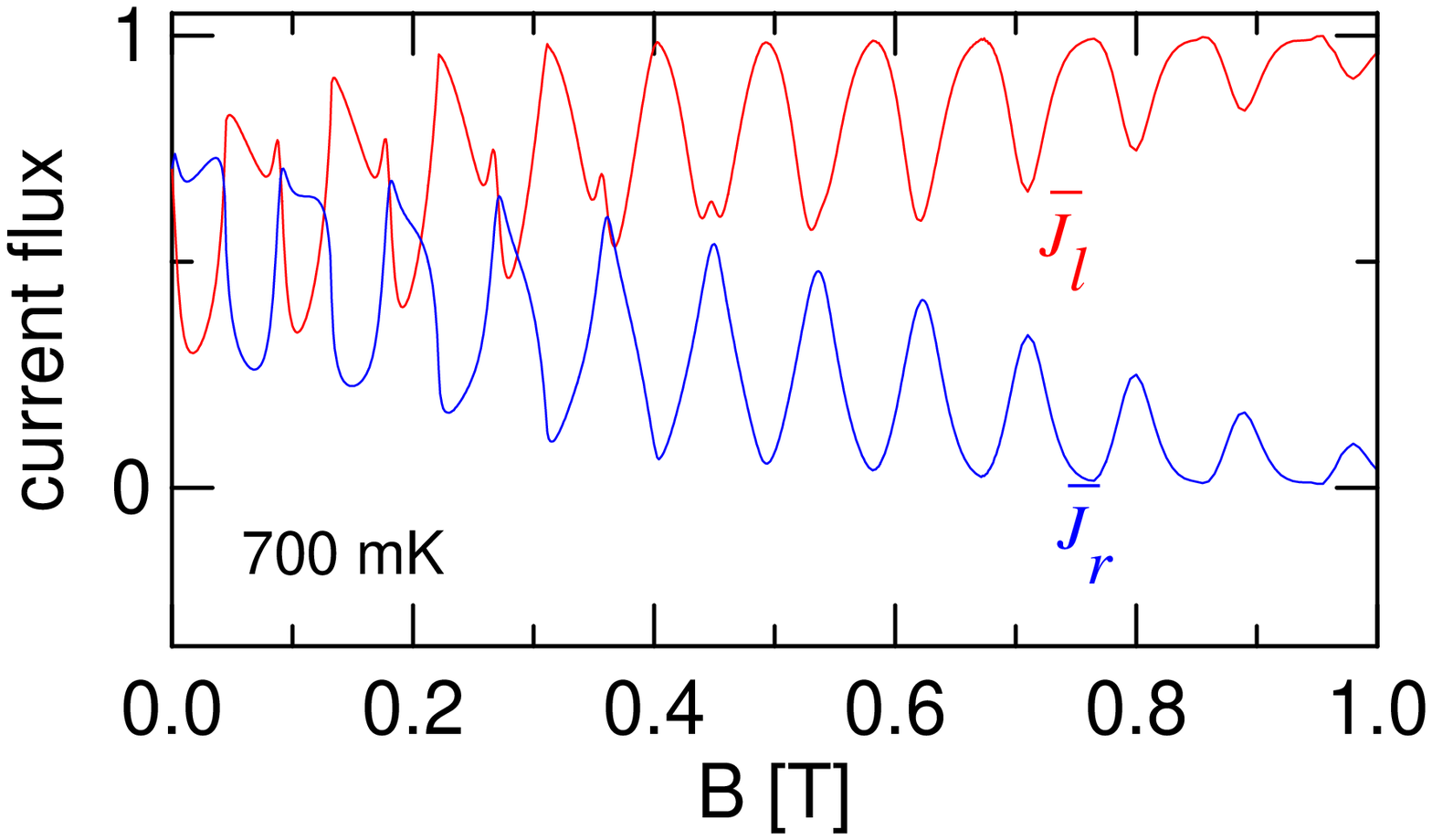}& f)\\
    \end{tabular}
    \caption{(a,c,e) Transfer probabilities to the left $\overline{T_l}$ and right output lead $\overline{T_r}$ as well as their sum $\overline{T}$
    averaged over the thermally widened transport window for     for $k_F=0.0667$ nm$^{-1}$.
        (b,d,f) Normalized current fluxes through the left and right arm of the ring.
      The results are presented for the temperatures $\tau=115$ mK (a,b) and $350$ mK (c,d) and 700 mK (e,f).}
    \label{temp}
    \end{figure*}

    \subsection{Finite temperature effect}
    At high magnetic field the interference conditions leading to anomalous injection of the current to the
    right arm of the ring appear for narrow $k$ intervals.
    The conductance measurements are performed in finite temperatures of the order of 100 mK,\cite{strambini,ferrier}
    for which a transport window of a finite width is opened  near the Fermi level.
    In order to study stability of these anomalous transport conditions in finite temperatures
    we performed calculations for averaged transfer probabilities according to  Eq. (\ref{tildet}).

    For the temperature $\tau=115$ mK the weight function $-\frac{\partial f}{\partial E}$ calculated for Fermi wave vector\cite{kkf} $k_F=0.0667$ nm$^{-1}$
    is nearly a Gaussian function of $k$ centered at $k_F$ with half width
    $\Delta k=4.5\times 10^{-4}$ nm$^{-1}$ for $B=0$ and $\Delta k=5.2\times 10^{-4}$ nm$^{-1}$  for $B=0.8$ T.
In the $B\rightarrow\infty$ limit the energy tends to the lowest Landau level for any wave vector  $E(k)\rightarrow\hbar\omega_c/2$, hence the widening of the $k$ window for a given thermal energy $k_b\tau$ at higher $B$.
    For $\tau=350$ mK (700 mK) the corresponding half widths are $\Delta k=1.4\times 10^{-3}$ nm$^{-1}$ ($3\times 10^{-3}$ nm$^{-1}$)
    and $\Delta k=1.6\times 10^{-3}$ nm$^{-1}$ ($3.5\times 10^{-3}$ nm$^{-1}$), for $B=0$ and $0.8$ T, respectively.


    Fig. \ref{temp} shows the transfer probabilities and normalized current fluxes
    for Fermi wave vector fixed at $k_F=0.0667$ nm$^{-1}$ and three values of the temperature
    (results for 0 K were given in Fig. \ref{wb667}).
    In finite temperature the dips and peaks of the transfer probabilities are transformed into smooth extrema
    of reduced amplitude which eventually disappear at high magnetic field.
    The attenuation of the Aharonov-Bohm oscillations of the transfer probabilities for non-zero temperatures at higher $B$ is in agreement with
    the  results of previous wave packet simulations\cite{time,epl,poniedzialek}, in which the averaging of the transfer probabilities with $k$ are embedded in the initial
    condition. The attenuation was also observed in the experimental data of  Ref. [\onlinecite{strambini}].
    Results of Fig. \ref{temp} indicate that the oscillations
    of the direction of the current circulation around the ring, which determine the orientation of the generated magnetic dipole moment,
     are more thermally stable than the oscillations of the transfer probabilities, which determine the conductance.

    \begin{figure*}[ht!]
    \begin{tabular}{llll}
     a) &  \epsfxsize=60mm \epsfbox[31 98 555 391] {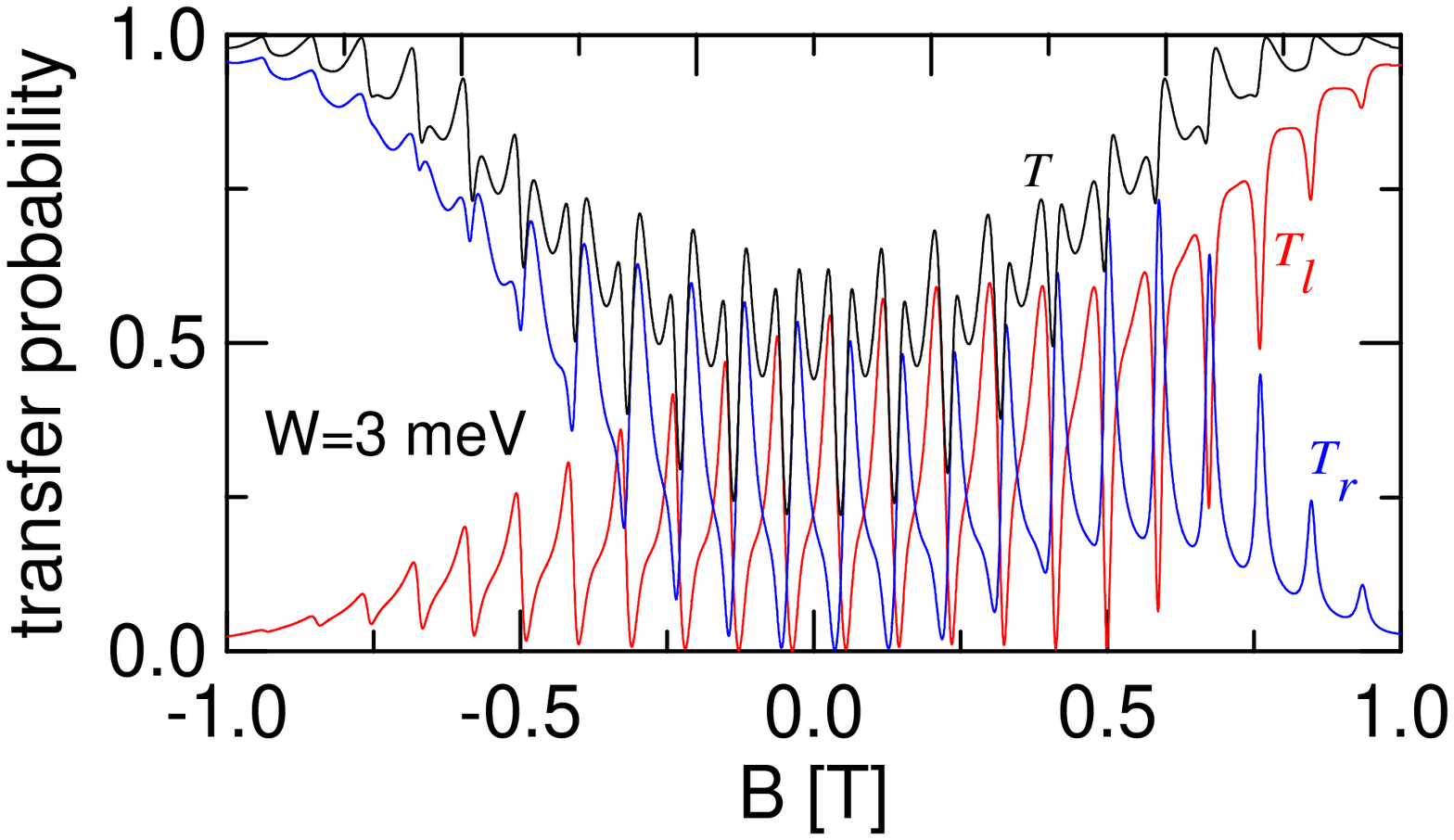} & \epsfxsize=60mm \epsfbox[31 98 555 391] {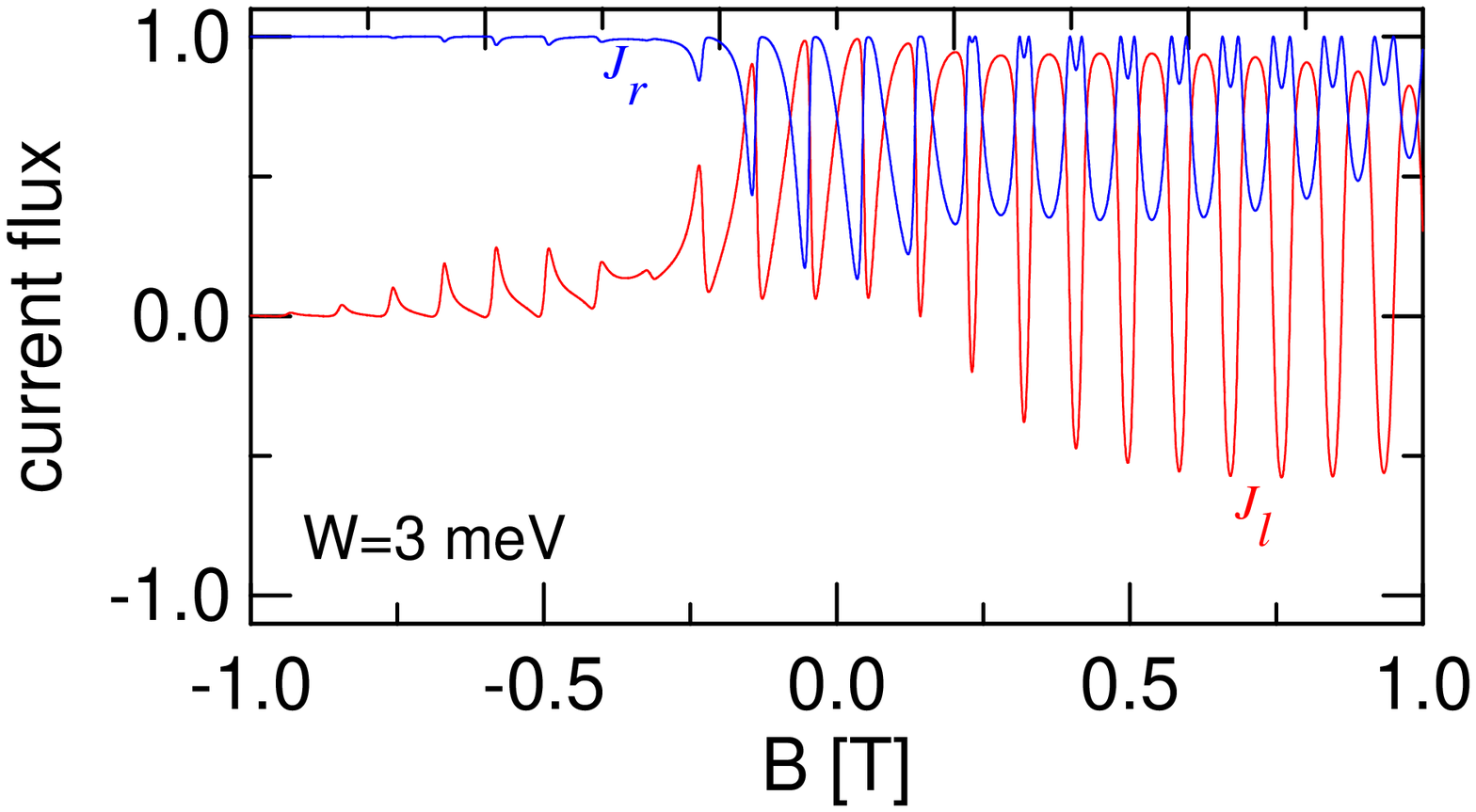}& b) \\
      c) &  \epsfxsize=60mm \epsfbox[31 98 555 391] {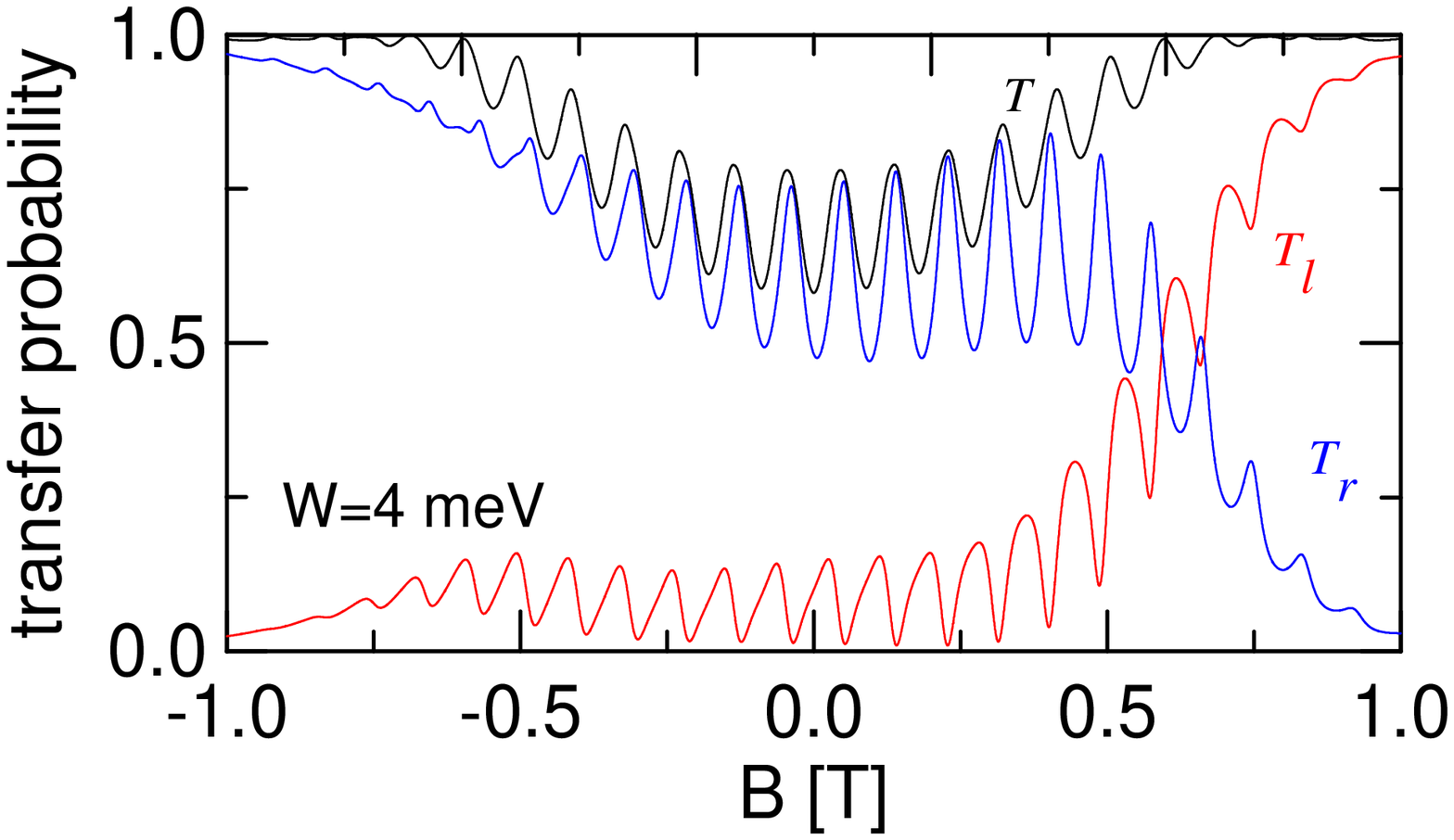}  & \epsfxsize=60mm \epsfbox[31 98 555 391] {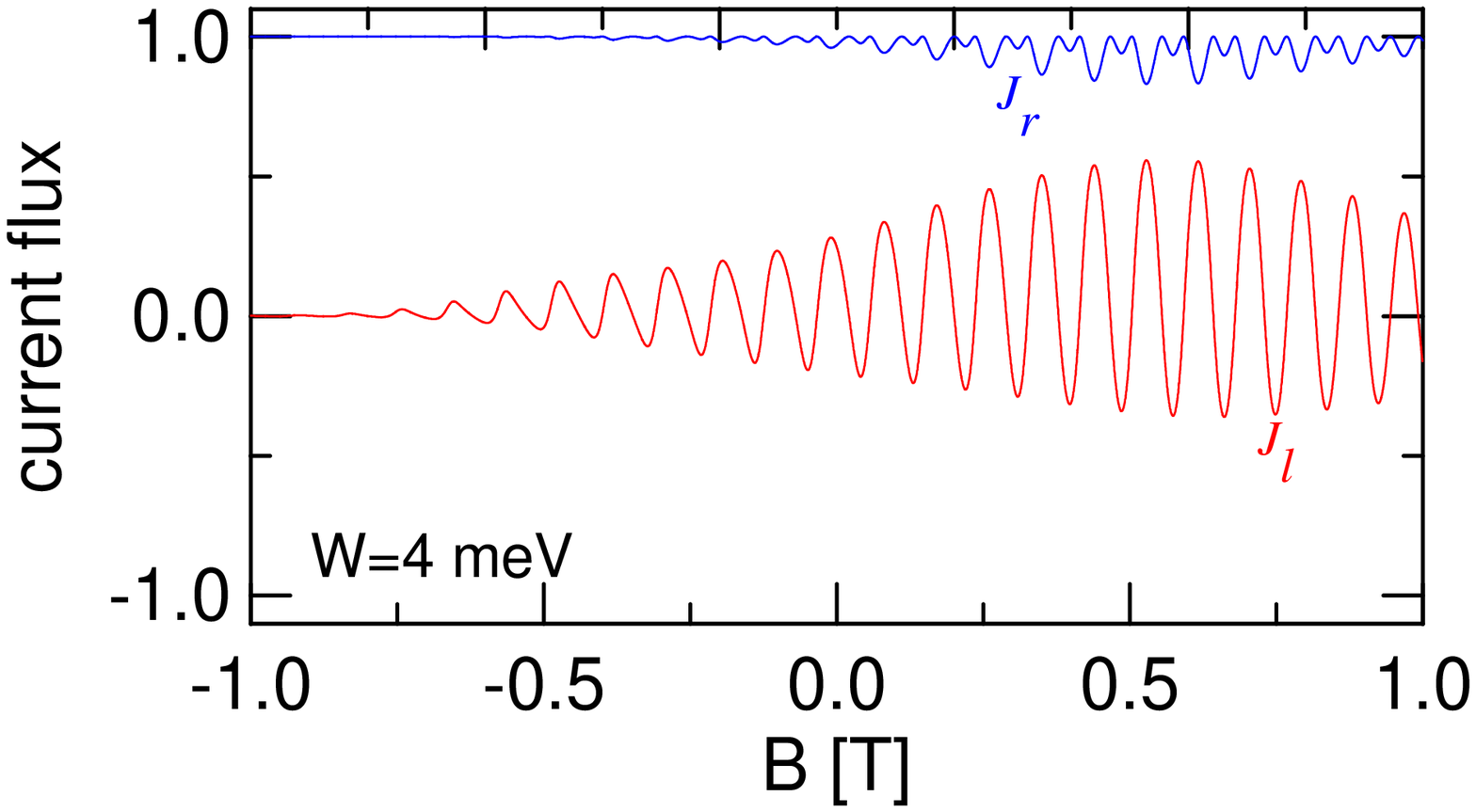}& d)\\
        e) &  \epsfxsize=60mm \epsfbox[31 98 555 391] {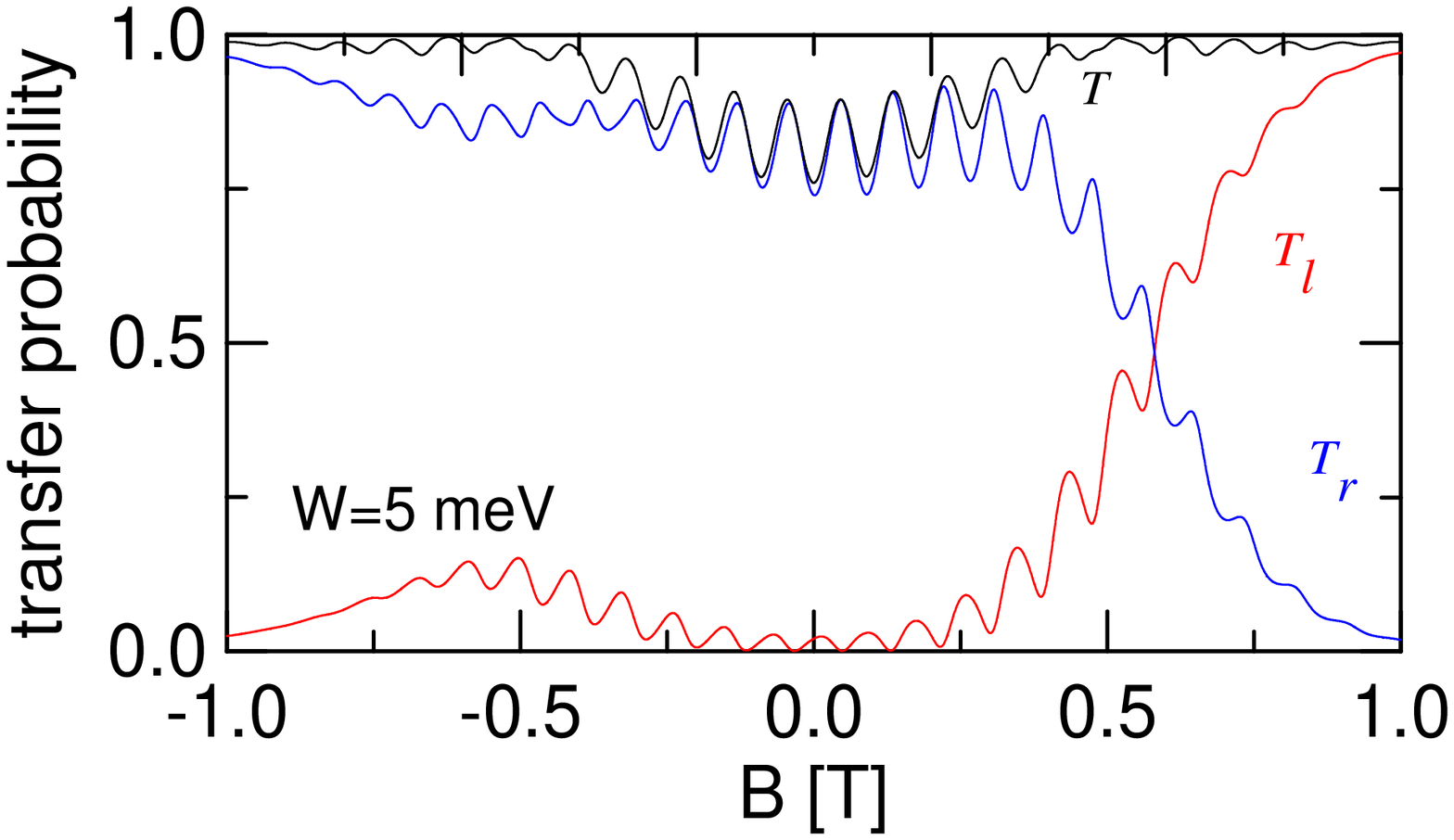}  & \epsfxsize=60mm \epsfbox[31 98 555 391] {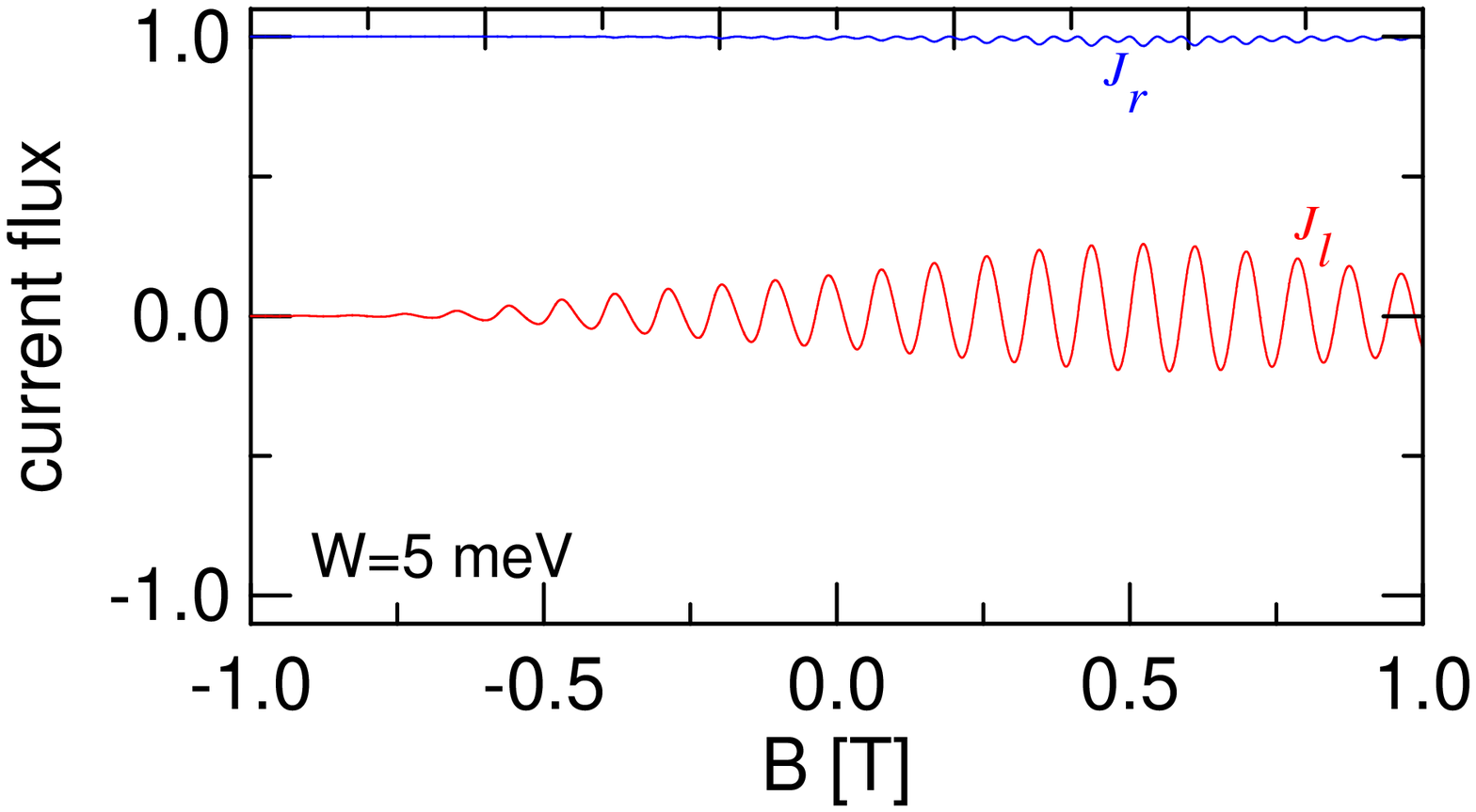}& f)\\
            g) &  \epsfxsize=60mm \epsfbox[31 98 555 391] {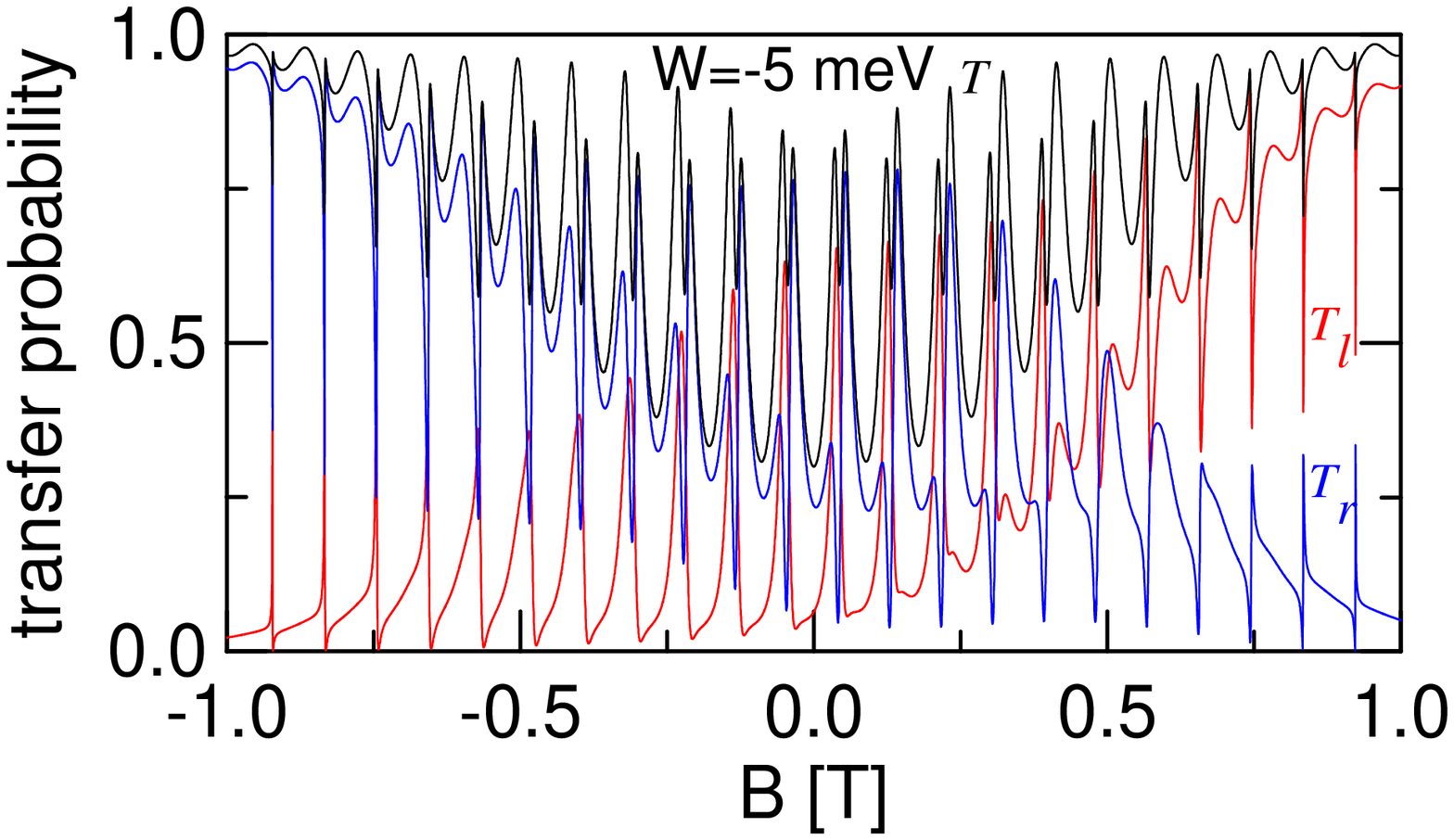}  & \epsfxsize=60mm \epsfbox[31 98 555 391] {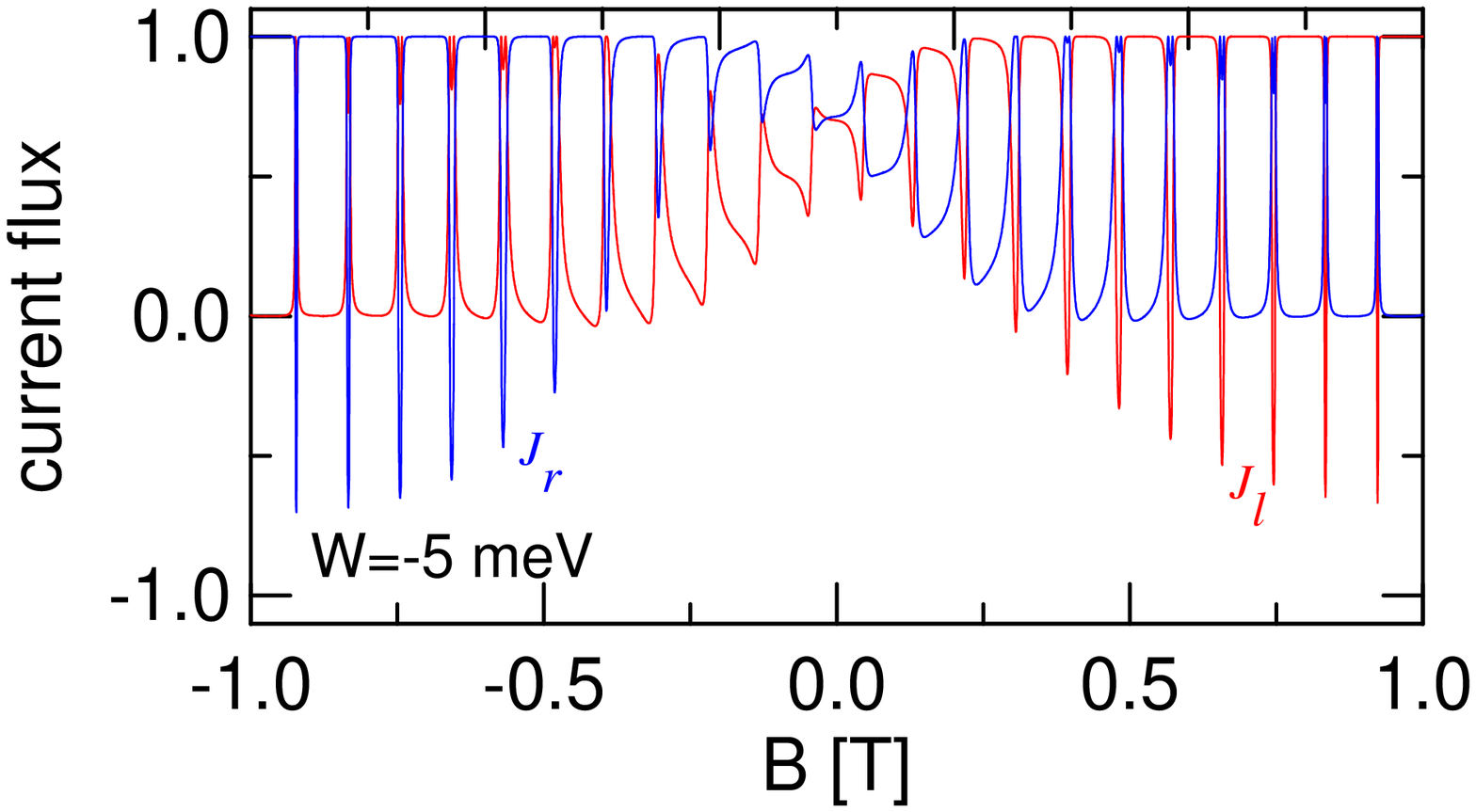}& h)\\
    \end{tabular}
    \caption{The transfer probabilities and normalized current fluxes for a repulsive potential defect of height  3 meV (a,b), 4 meV (c,d) and 5 meV (e,f).
    Plots (g,h) correspond to an attractive defect of depth $-5$ meV.}
    \label{domi}
    \end{figure*}

    \subsection{Ring with a perturbed potential}

    The experimental results (Fig. 1 of Ref. [\onlinecite{strambini}]) indicate a significant anisotropy of the potential landscape
    within the ring since already at $B=0$ the conductance of one of the output leads largely exceeds the other.
    The appearance of peaks of $T_r$ at high $B>0$ that we discussed above were associated with specific
    interference conditions for which the electron wave function at the junction to the left output lead
    possessed a minimum at the axis of the lead (see Fig. 10 for $B=0.3625$ T for instance).
    A question which seems natural is whether such interference conditions
    are still possible for a quantum ring containing a potential defect.

    In order to answer this question we considered a perturbation introduced by Gaussian potential $V_d=W \exp(-[(x-x_c)^2+(y-y_c)^2]/R_d^2]$,
    centered in point $x_c=-104.8$ nm $y_c=179.5$ nm in the left arm just in between the input and left output leads. The size of the defect
    is assumed $R_d=30$ nm. The results for the transfer probabilities and current fluxes are displayed in Fig. \ref{domi}.

    For  $W=-5$ meV the impurity introduces a potential cavity which mainly shifts the phase of the
    wave function passing through the left arm [Fig. \ref{domi}(g,h)]. We observe no pronounced effect for the qualitative features of the transfer
    probabilities at high magnetic field as compared to a clean ring $W=0$ case (cf. Fig. 8).

    A potential barrier that is introduced for $W>0$ hampers the electron transfer through
    the left arm. For $W\ge 4$ meV the transfer probabilities to the left and right output lead become distinctly different near $B=0$ [Fig. \ref{domi}(c,e)], the
    amplitude of the Aharonov-Bohm oscillation is significantly reduced and the peak / dip structures disappear in the high field limit.

    Results of Fig. \ref{domi} for $W=4$ and $5$ meV resemble the measured conductance.\cite{strambini}
    Near $B=0$ the electron transfer goes mainly to the right lead. $T_l$ exceeds $T_r$ only for $B>0.5$ T.
    Note, that also for $B>0.5$ T the current flux through the right arm of the ring largely exceeds the one through the left arm
    [Fig. \ref{domi}(d,f)].
    The dominant electron trajectory for this transport conditions was indicated in Ref. [\onlinecite{poniedzialek}] using  wave packet simulations.

     We conclude that the presence of a repulsive potential defect induces not only
    the asymmetry of the transfer at $B=0$ and a weak amplitude of the Aharonov-Bohm oscillation
    but also the  absence of $T_r$ peaks at high $B$. For a strongly asymmetric potential the peaks of $T_r$ disappear also
    in zero temperature.

    \begin{figure}[ht!]
    \begin{tabular}{ll}
     a) &  \epsfxsize=80mm \epsfbox[31 138 577 433] {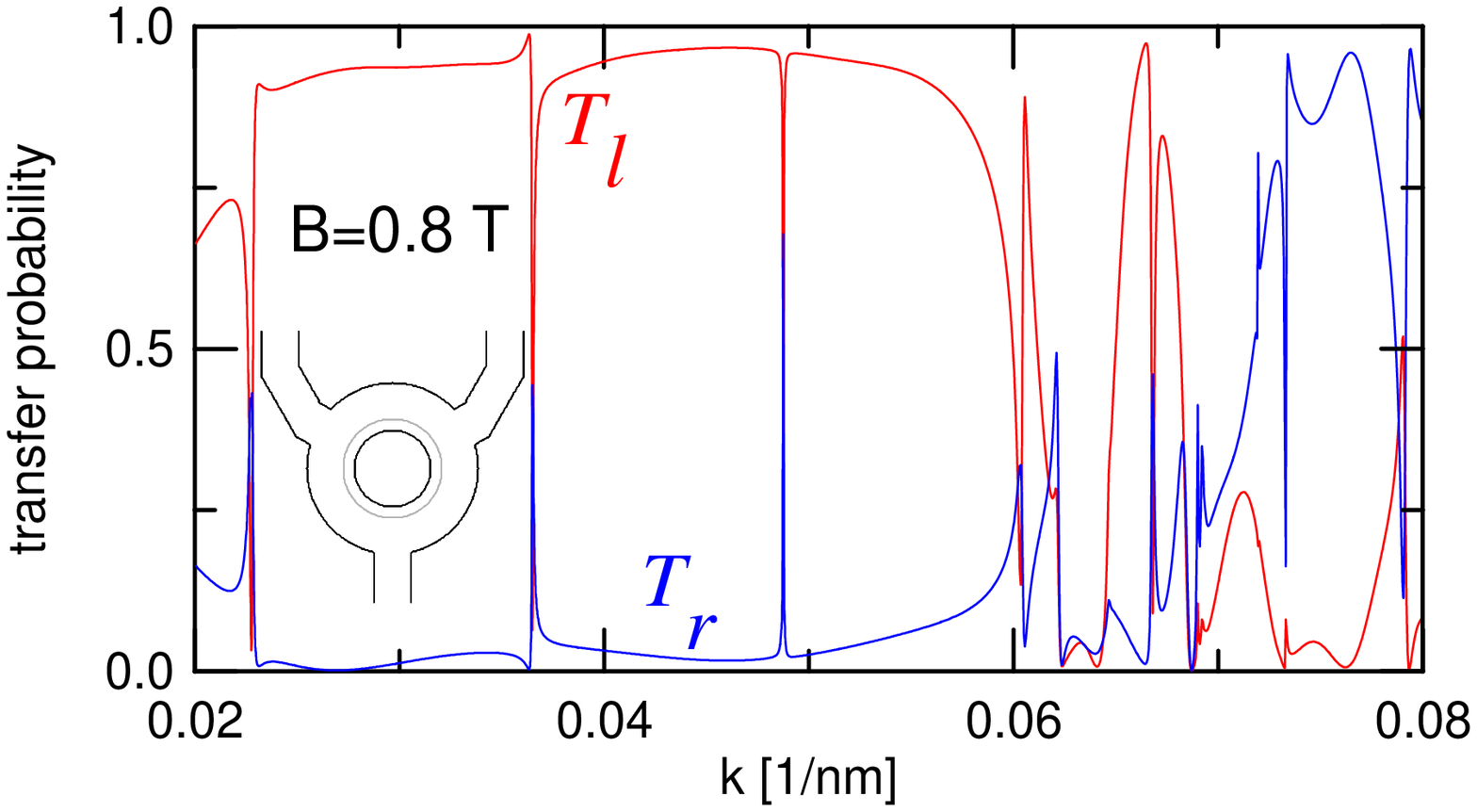}  \\
     b) &  \epsfxsize=80mm \epsfbox[31 138 577 433] {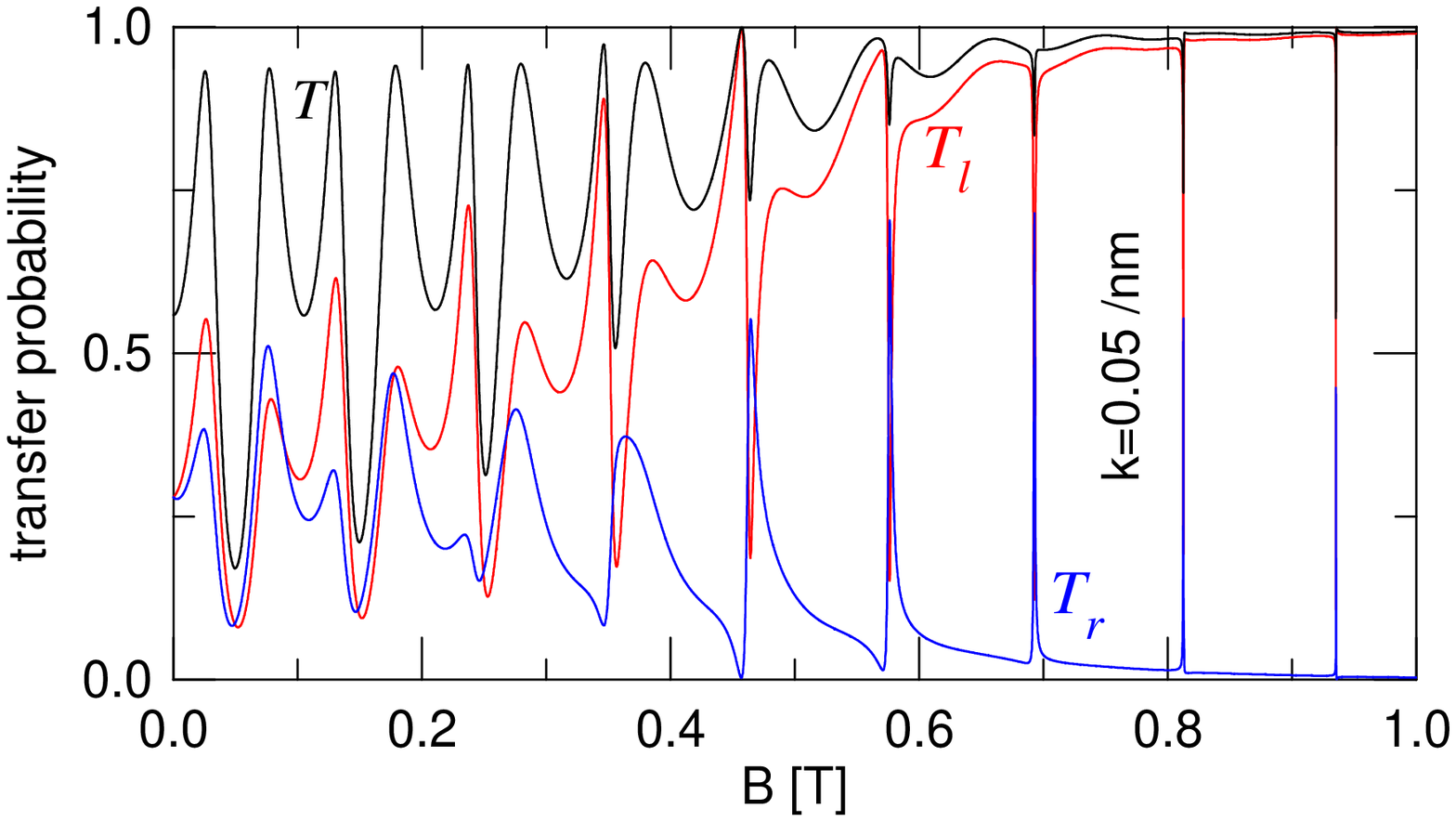}  \\
      c) &  \epsfxsize=80mm \epsfbox[31 138 577 433] {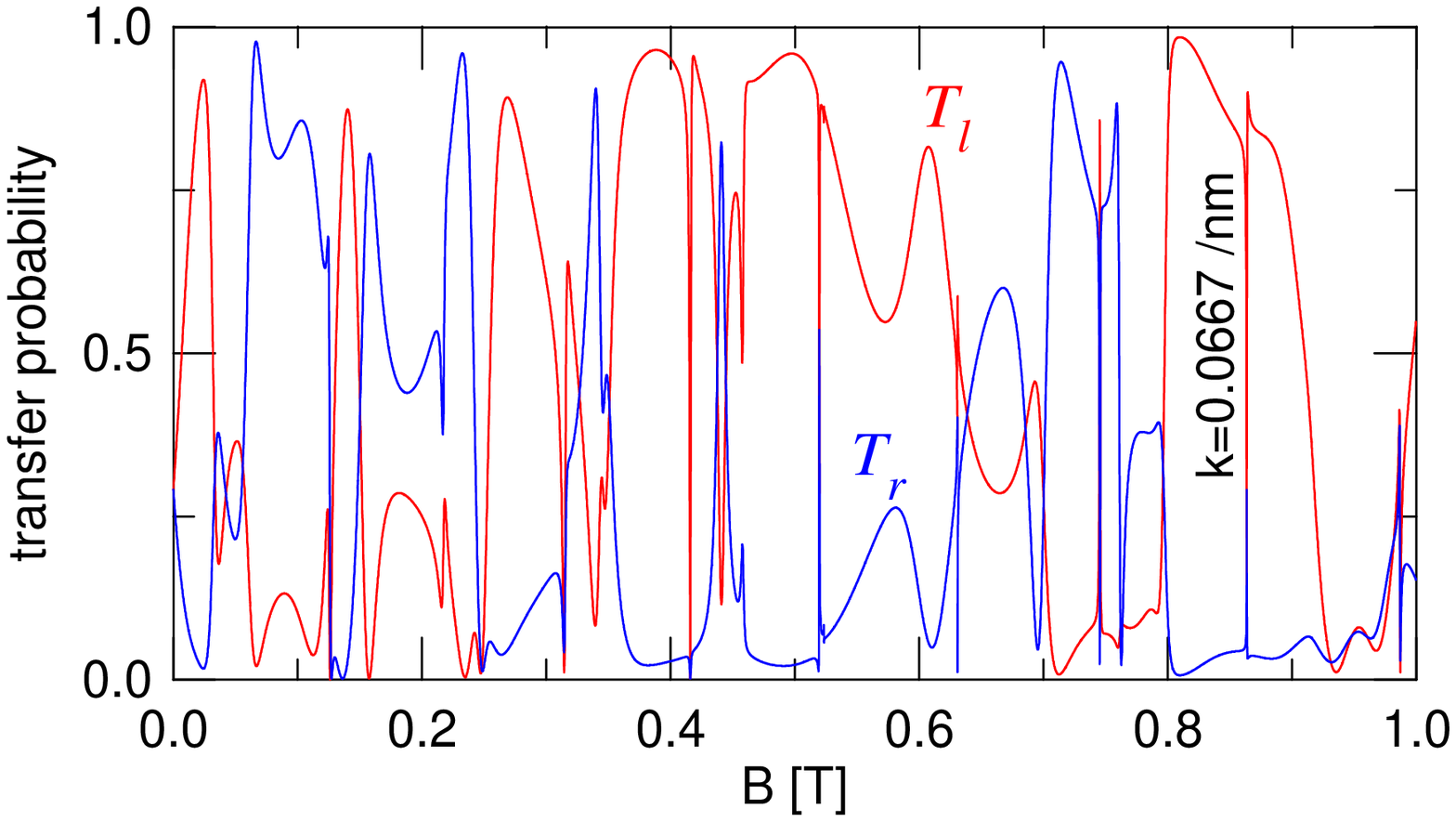} \\
    \end{tabular}
    \caption{(a) Wave vector resolved transfer probabilities to the left and right output leads for
    the inner ring radius decreased from 88 nm (grey circle in the inset) to 68 nm (the black circle inside the grey one).
    Results for the ring channel width equal to the with of the lead channels were presented in Fig. \ref{wk}(b).
    Transfer probabilities as functions of $B$ are plotted for $k=0.05$$^{-1}$ and $k=0.0667$ nm$^{-1}$ in (b) and (c), respectively.}
    \label{wide}
    \end{figure}

    \subsection{Ring of an increased channel width}
    The above results were obtained for the width of the channel within the ring fitted to the width of terminals.
    For an increased width of the ring channel the electron coming of the lowest subband of the input lead
    may possess enough energy to occupy locally -- i.e. within the ring -- the second subband. A local scattering to the second subband may influence the mechanism
    of the electron transfer through the system. In order to study this point we decreased the inner radius of the ring
    from 88 to 68 nm (see the inset to Fig. \ref{wide}).

    The $k$-resolved transfer probabilities are plotted for $B=0.8$ T in Fig. \ref{wide}. For $k<0.06$ nm we find
    similar results to the ones presented above: the transfer goes to the left output lead for nearly each value $k$.
    For $k>0.06$ nm$^{-1}$ the scattering to the second subband of the channel becomes allowed and one
    observes a non-regular dependence of the transfer probabilities, with $T_r$ exceeding $T_l$
    on some intervals. For $k=0.05$ nm$^{-1}$  [Fig. \ref{wide}(b)] the transfer probabilities change with the magnetic field in the same
    manner as for the ring of smaller width. Very different results are obtained for $k=0.0667$ nm $^{-1}$  [Fig. \ref{wide}(c)].
    Not a sign of  periodicity can be noticed. The results seem chaotic
    with no clear signature of the Lorentz force effect. The results for $k>0.06$ nm$^{-1}$  resemble  rather the transport through a chaotic cavity
    (quantum billiard\cite{cb}) than through a quantum ring.

    \section{Summary and Conclusions}
    We have discussed the role of magnetic forces in stationary electron flow through  a three-terminal quantum
    ring as obtained for Hamiltonian eigenstates in a single subband transport regime.
    We have shown that in most cases at high magnetic field the transport seems governed
    by the magnetic forces: the entire current is injected into the left ($B>0$) arm of the ring
    and then ejected to the left output lead, with the transfer probability that tends to 100\% at high magnetic fields.
    Exception to this rule are found only for narrow windows of magnetic fields for which
    interference conditions within the ring lead to formation of wave functions which are weakly coupled to
    the left output channel. This form of interference is associated with anticlockwise circulation
    of the current within the ring and with an appearance of narrow peaks of transfer probabilities to the right output lead.
    The anticlockwise circulation is anomalous from the  point of view of the direction of classical magnetic forces since the
    current is injected into the right and not the left arm of the ring.
    The sharp peaks of the transfer probability to the right output lead that are found for high $B$ disappear
    in finite temperatures for which the Aharonov-Bohm oscillations of conductance are eventually attenuated.
    Oscillations of the current circulation turn out to be more resistant to the thermal widening of the Fermi level than the transfer probabilities.
    We have demonstrated that
    the  imbalance of the transfer probabilities at $B=0$ as well as the reduction of the conductance oscillations
    that are introduced by a scattering center within the ring are associated with removal of the interference
    conditions leading to appearance of the peaks of $T_r$ at high magnetic field.
    We have considered the ring with a width larger than the width of the channels. We demonstrated
    that the for wave vectors which allow for appearance of local scattering to an excited subband within the ring  channel --
    the results for conductance become chaotic in function of $B$ without a clear signature of either the Lorentz force effect
    or the Aharonov-Bohm oscillations.

    {\bf Acknowledgements} This work was supported by the
    AGH UST project 11.11.220.01 "Basic and applied research in nuclear
    and solid state physics". Calculations were performed in
    ACK\---CY\-F\-RO\-NET\---AGH on the RackServer Zeus.


\begin{thebibliography}{00}
\bibitem{kkf} In the present calculation the relation between the wave vectors and energy, $k_F$ and $E_F$ in particular are given by Eq. (\ref{new2}).
\end{thebibliography}
    \end{document}